\documentclass[aip,preprint,groupedaddress,11pt]{revtex4}
\usepackage{natbib}
\usepackage{graphics}
\usepackage{epsfig}
\usepackage{graphicx}
\usepackage{color}
\usepackage{amsmath}
\usepackage{amssymb}
\usepackage{ulem}
\newcommand*{\beq}{\begin{eqnarray}}
\newcommand*{\eeq}{\end{eqnarray}}
\newcommand*{\bea}{\begin{eqnarray}}
\newcommand*{\eea}{\end{eqnarray}}

\newcommand \qvec{{\bf q}}

\newcommand \hR{\hat{R}}
\newcommand \hr{\hat{r}}

\newcommand\rvec{{\bf r}}
\newcommand\bs{{\bf s}}
\newcommand\bb{{\bf b}}

\newcommand{\1}[1]{{\color{black}{#1}}}

\def\simge{\mathrel{%
       \rlap{\raise 0.511ex \hbox{$>$}}{\lower 0.511ex \hbox{$\sim$}}}}
\def\simle{\mathrel{
       \rlap{\raise 0.511ex \hbox{$<$}}{\lower 0.511ex \hbox{$\sim$}}}}

\begin{document}

\title{Consistent and transferable coarse-grained model for semidilute polymer solutions in good solvent}

\author{Giuseppe D'Adamo}
\affiliation{Physics Department, University of L'Aquila, Via Vetoio, 67100 L'Aquila, Italy}
\author{Andrea Pelissetto}
\affiliation{Physics Department, ``Sapienza" University of Rome, and INFN, Sezione di Roma I, P.le A.Moro 2, 00185 Rome, Italy}
\author{Carlo Pierleoni}
\email{carlo.pierleoni@aquila.infn.it}
\affiliation{Physics Department, University of L'Aquila and CNISM, UdR L'Aquila, Via Vetoio, 67100 L'Aquila, Italy}

\date{\today}

\begin{abstract}

We present a coarse-grained model for linear polymers with a tunable number 
of effective atoms (blobs) per chain interacting by intra- and inter-molecular potentials
 obtained at zero density. We show how this model is able to accurately reproduce the universal properties of the underlying solution 
of athermal linear chains at various levels of coarse-graining and in a range 
of chain densities which can be widened by increasing the spatial resolution 
of the multiblob representation, i.e., the number of blobs per chain. 
The present model is unique in its ability to quantitatively predict 
thermodynamic and \1{large scale} structural properties of polymer solutions deep in the 
semidilute regime with a very limited computational effort, overcoming most 
of the problems related to the simulations of semidilute polymer 
solutions in good solvent conditions. 

\end{abstract}


\maketitle

\section{Introduction}
The last decade has witnessed a considerable effort in developing 
coarse-grained (CG) models to bridge the length-scale gap between the 
microscopic scale and the meso(macro)-scopic scale typical of soft-matter 
and biological systems. Two general strategies have been attempted: 
the structure-based route in which CG models are tuned to 
reproduce some structural properties at the local scale and the 
thermodynamic-based route in which the models are required to reproduce 
solvation free energies. Also mixed strategies have been developed and 
applied to a large variety of physical systems. An overview of methods and 
recent applications can be found in Refs. \cite{Voth2009,PCCP2009,SM2009,FarDisc2010}. 

The structure-based route groups a number of atoms into effective ``particles'' 
and assumes state-dependent pair interactions between them. These potentials are 
derived from the local structure of the atomic-level system, using the 
Iterative Boltzmann Inversion (IBI) 
\cite{Carbone:2008p2254,Fritz:2009p1721,Peter:2009p1734,Tschop98}, 
the Inverse Monte Carlo (IMC) \cite{Lyubartsev:2009p2497} or by liquid-state 
theory techniques \cite{Hansen:2005p292,Clark:2010p2483}. The state dependence of the 
effective pair potentials hides the underlying many-body character and 
entropic content of the effective interactions and poses the question of 
their transferability. Indeed, if the effective potentials are 
not transferable, setting up the CG model requires deriving the effective 
interactions for all thermodynamic states of interest,
seriously limiting the benefits of the coarse-graining strategy.
Generally speaking, the problem of transferability remains unsolved.

In the specific case of linear polymers in solutions, 
structure-based CG models, where a single coil is mapped onto a single soft blob, have been introduced
quite some time ago \cite{Flory:1950p2484,Grosberg:1982p2265,Dautenhahn:1994p2250,%
Murat:1998p1980,Bolhuis:2001p268,D'Adamo2012,Pelissetto:2005p296}. For two isolated coils in athermal solvent 
the effective interaction between the centers of mass of the coils is roughly 
Gaussian, of the order of $2k_BT$ at overlap, and with a range of the 
order of the coil radius of gyration. This single-blob model is only
accurate in the dilute regime, in which chain overlaps can be neglected.
Extension of the single-blob CG model to the semidilute regime can be obtained by allowing 
the use of density dependent pair potentials\cite{Bolhuis:2001p268} which can be determined from
the radial distribution function between the centers of mass of two coils computed 
from a full monomer simulation at finite density.
This strategy however suffers for the transferability problem since setting up the coarse-grained model requires simulations of the 
underlying full monomer system at all densities of interest.
Another limitation of the single-blob model with density dependent potential is in  
representing non-homogeneous systems since the interaction should depend on the local density which is 
not known beforehand and some kind of self-consistent procedure should be 
developed. Furthermore, representing polymers as soft spherically symmetric 
particles is not always appropriate. For instance, in studying polymers 
adsorbed on surfaces, like polymer brushes or polymer-coated colloids, 
it is clear that the anchorage to the surface breaks the rotational symmetry 
of the chains, an effect that must be taken into account in any accurate 
coarse-grained model\cite{Coluzza2008}. A further example is in modelling solutions of 
A-B block copolymers which cannot be represented as soft particles interacting 
by a spherically symmetric pair potential \cite{Pierleoni:2006p159,HansenJP:2006p2248,Sambriski:2007p781,Capone2009,Gross:2010p2528,Capone2010,Capone2011}.

In a semidilute solution of $N$ linear chains of $L$ 
monomers in a volume $V$, chain density, $c=N/V$, is larger than the
overlap density $c^*=3/4\pi \hat{R}_g^3$ ($\hat{R}_g=bL^{\nu}$ is 
the isolated coil radius of gyration, $b$ the monomer size and $\nu$ 
the scaling exponent)
while the monomer density is still very small, a condition that can always be
satisfied if chains are long enough \cite{Doi}.
An accurate description of the thermodynamic and \1{large-scale} structural behavior of 
polymers in these conditions can in principle be obtained by using CG chains, in 
which a number $m$ of the original monomers are grouped in one effective 
monomer (blob) to map the original chain of $L=nm$ monomers onto a chain 
of $n$ effective blobs (multiblob model). If the level of coarse graining, 
i.e. the number $n$ of blobs per chain, is such that the blob density $c_b=nc$ 
is below the overlap blob density $c_b^*=3/4\pi \hat{r}_g^3$ 
($\hat{r}_g\sim m^{\nu}$ is the zero-density
radius of gyration of the blob), then it is expected that zero-density  
potentials between blobs or different chains can be safely used.
These ideas were discussed in 
Refs.~\cite{Pierleoni:2007p193,Pelissetto:2009p287,Vettorel:2010p1733} and recently reviewed in 
Ref.~\cite{DAdamoG:2011p2477}. 
Although in principle very appealing, the problem with this approach is how to
obtain the intramolecular potential. Indeed, this inherently many-body 
potential is in principle of increasing complexity when increasing the number
of blobs per chain, and since blobs are tethered together in some fixed
topology, a zero-density expansion cannot be invoked to decompose it into 
a sum of two-body, three-body, etc. terms.
In Ref. \cite{DAdamoG:2011p2477} a coarse-grained model for good-solvent conditions with four effective monomers per chain (tetramer)
was developed and throughly studied. Such a model, set up at zero density, was found to be accurate up to \1{reduced} polymer density 
\1{$\Phi=c/c^*\simeq 2$} supporting the multiblob ideas. 

In this paper we present the extension of the \textit{tetramer} model to chains with an arbitrary number of blobs, 
an extension needed to explore the semidilute regime at high polymer concentration. We show that a careful parametrization
of the $n$-body interactions up to $n=4$ is enough to obtain a very accurate CG potential field fully transferable both with the 
number of blobs per chain and with the density of chains\1{, provided that the length of the effective chains is chosen to ensure that the reduced blob density remains always low}. 
The paper is organized as follows. In section II we will provide a theoretical framework on which any accurate coarse-grained model should be based. In the following section III we present our multiblob model and report results for systems of chains of varying length both at zero density (section III.a) and at finite density in the semidilute regime (section III.b). In section IV we collect our conclusion and perspectives. Finally an appendix reports the tetramer potentials as obtained by the IBI and their explicit parametrization.

\section{Theoretical framework}\label{sec:theory}
A crucial requirement of any accurate \1{coarse-grained representation (CGR)} representation of the
underlying full-monomer (FM) chain model is to preserve the value of the 
radius of gyration for all values of \1{the number of blobs} $n$.
Let us consider a chain of $L$ monomers mapped onto a chain of $n$ blobs, 
each representing the center of mass of a subchain of $m$ monomers. The 
fundamental relation  
$R_g^2=R_{g,b}^2(n)+r_g^2(n)$ among the chain radius of gyration $R_g$, 
the radius of gyration $R_{g,b}(n)$ of the chain of $n$ blobs, and the 
mean blob radius of gyration $r_g(n)$, holds for any single configuration and 
therefore on average. Here $r_g^2(n)$ is the average of the square radius of gyration of all blobs in the chain
\beq
r_g^2(n)=\frac{1}{n}\sum_{k=1}^n r_{g,k}^2(n),
\label{eq:rg2^}
\eeq
and 
\beq
r_{g,k}^2(n)=\frac{1}{2m^2}\sum_{i,j=m(k-1)+1}^{mk} (\rvec_i-\rvec_j)^2,
\eeq
where $\rvec_i$ is the position of the $i$-th monomer in the chain. If $\bs_i$ is the center of mass of the $i$-th blob in the chain,
\beq
\bs_i=\frac{1}{m}\sum_{j=m(i-1)+1}^{mi}\rvec_i,
\eeq
$R_{g,b}(n)$ is defined by
\beq
R_{g,b}^2(n)=\frac{1}{2n^2}\sum_{i,j=1}^n (\bs_i-\bs_j)^2.
\label{eq:Rgb}
\eeq
In the scaling limit, it has been found that \cite{DAdamoG:2011p2477}
\bea
&&\frac{\hR_{g,b}^2(n)}{\hR_g^2}= \left(1-\frac{k^2}{n^{2\nu}}\right), \label{equation:eq1}\\
&&k^2=\frac{\hr_g^2(n) n^{2\nu}}{\hR_g^2}= \left(1.03-\frac{0.04}{n}\right)^2,\label{equation:k2}
\eea
where $\nu=0.587597(7)$ \cite{Clisby:2010p2249}. Here and in the following we 
will use a hat to indicate zero-density averages. For $n\gg 1$, 
$\hR_{g,b}(n)\simeq \hR_g$ and $\hR^2_{g,b}(n)/\hr_g^2(n)=0.94 n^{2\nu}$.

Any consistent multiblob (MB) model 
must obey these relations when varying $n$.  
A proper definition of $\hR_g$, independent on $n$, including the prefactor, is crucial since 
chain structural properties are universal only if distances are expressed in terms 
of rescaled distances $\rho=r/\hR_g$. 
Moreover, at finite density results from different 
values of $n$ should be compared at the same value of the polymer volume 
fraction $\Phi=c/c^*=(4\pi \hat{R}_g^3)c/3$. Changing the 
definition of $\hR_g$ with $n$, changes the definition of both $\rho$ and $\Phi$ and the 
comparison among models with different $n$ becomes meaningless. 

In Ref.~\cite{DAdamoG:2011p2477} a MB model with four blobs per chain 
(tetramer) was developed and throughly studied. 
Central potentials between first, second and third neighbors along the chain,
as well as a bending and torsional angle potentials were determined by IBI to
reproduce the pair distances and angular distributions of \1{an isolated} FM chain in 
the scaling limit, mapped onto four blobs \cite{DAdamoG:2011p2477}. Moreover, 
an intermolecular Gaussian potential between any pair of blobs of different 
tetramers was assumed and optimized to reproduce the radial distribution 
function between the centers of mass of two \1{isolated} chains in the scaling limit 
\cite{Pelissetto:2005p296}.
The model perfectly reproduces the structure of the chains and 
provides accurate results for the thermodynamics of the solution 
up to $\Phi\simeq 2$. 
The non-uniform (universal) angular distributions of the tetramer are induced by the 
specific mapping of subchains onto their centers of mass, a procedure which 
introduces an explicit angular correlation along the chain even for ideal 
chains \cite{Laso:1991p1417}. Failure in reproducing the correct angular 
distributions is at the basis of the inconsistent behavior observed for 
simpler multiblob models 
\cite{Pierleoni:2007p193,Pelissetto:2009p287,DAdamoG:2011p2477}.
Note that while central potentials are two-body interactions, 
bending and torsional angle 
potentials represent genuine three- and four-body interactions, respectively. 
Interestingly \cite{DAdamoG:2011p2477}, a hierarchical order in the intensity 
of the various potentials is observed, the strongest and most relevant being 
the central first-neighbor interaction (bonding), followed by the second- and third-neighbor 
central interactions, and finally by the bending and the torsional 
angle potentials.  Therefore, correlations along the chain decrease with the 
chemical distance and many-body effects are smaller than two-body ones, 
although of different nature. This observation suggests to use tetramers as 
building blocks of a resolution-invariant multiblob model.

\section{Tetramer-based multiblob models and results}\label{sec:model}

\subsection{Zero-density transferability}
A first tetramer-based multiblob model (4MB-1) is obtained by using the tetramer
potentials of Ref.~\cite{DAdamoG:2011p2477} reported in the Appendix. 
In model 4MB-1 one considers six different intramolecular potentials, 
each of them depending on a single scalar variable. 
We use four bonding pair potentials: blobs $i$ and $j$ interact with a central 
potential $V_{ij}(\rho;4)$ with $\rho=\left|\bs_i-\bs_j\right|/\hat{R}_g$. 
Symmetry imposes $V_{1,2}(\rho;4)=V_{3,4}(\rho;4)$ and $V_{1,3}(\rho;4)=V_{2,4}(\rho;4)$ 
so that there are only four different central potentials. 
Then we use a three-body potential $V_b(\cos \beta_i)$ acting on the two equivalent bending angles and a four-body potential $V_t(\theta)$ acting on the torsional angle, where $\beta$ and $\theta$ are defined as
\bea
\cos \beta_i&=&-\frac{\bb_i\cdot \bb_{i+1}}{|\bb_i||\bb_{i+1}} \hskip 2cm (i=1,2),\\
\cos \theta&=&\frac{
    ({\bf b}_1 \times {\bf b}_{2}) \cdot 
    ({\bf b}_{2} \times {\bf b}_{3}) 
    }{ 
   |{\bf b}_1 \times {\bf b}_{2}|
   |{\bf b}_{2} \times {\bf b}_{3}| },
\eea
where ${\bf b}_i=\bs_{i+1}-\bs_i$ is the bond vector between blobs $i$ ad $i+1$. The intermolecular potential is instead represented by a unique blob-blob pair interaction $W(\rho;4)$ 
acting between any pair of blobs of different tetramers. 

When using this potential field to build chains of $n$ blobs, the blob radius of
gyration $\hr_g$ is taken as the length scale of the central potentials. In practice this means that when transferring the potentials from the model with 
4 blobs to the model with $n$ blobs, the following identification is adopted
\beq
V_{ij}(\rho;n)=V_{ij}(\rho_n;4), \\
W(\rho;n)=W(\rho_n;4)
\eeq
where,
according to Eq. (\ref{equation:k2}), $\rho_n$ is defined by
\beq
\rho_n=\rho\left(\frac{4}{n}\right)^{\nu}\frac{1.03-0.04/4}{1.03-0.04/n}.
\eeq
The angular
potentials are instead invariant under this scale transformation. 

This model, 4MB-1, is not found to be fully
satisfactory, as illustrated in table \ref{table:Rg}, where we report
results for the zero-density chain radius of gyration and the 
second dimensionless virial
coefficient for increasing $n$ for polymer chains in the 
scaling limit (FM) \cite{footnote}, the multiblob model (4MB-1),
and the modified multiblob model (4MB-2) still to be discussed. In particular,
in model 4MB-1, $\hR_{g,b}(n)/\hR_g$ becomes greater than 1 for 
increasing $n$, a fact clearly incompatible with the multiblob procedure 
[see Eq.~(\ref{equation:eq1})]. Also a systematic increase with $n$ in the second 
virial coefficient of 4MB-1 is observed which witnesses the inability of 
such model to be fully consistent at all resolutions.
\begin{table}
\caption{$\hR_{g,b}(n)/\hR_g$ and universal ratio
$A_2=B_2/\hR_g^3$ ($B_2$ is the second virial coefficient) 
as a function of $n$ for polymers obtained by full-monomer simulations
(FM-CGR), and for the original and modified 4MB models. The universal value for
polymers of the second virial coefficient combination is $A_2 = 5.500(3)$
\cite{Caracciolo:2006p587}.} 
\label{table:Rg}
\begin{tabular}{cccccc} 
\hline
& \multicolumn{3}{c}{$\hR_{g,b}(n)/\hR_g$}  & \multicolumn{2}{c}{$A_2=B_2~\hR_g^3$} \\
\hline
$n$  &  FM-CGR &  4MB-1& 4MB-2  & $A_2$(4MB-1)  & $A_2$(4MB-2) \\
\hline
 4  & 0.8921(1) & 0.8933(3) & 0.8903(2) & \1{5.597(1)} &  5.596(3)\\
10 & 0.9642(1) & 0.9812(3) & 0.9638(7) & 5.676(2) & 5.619(2)\\
20 & 0.9842(1) & 1.0060(6) & 0.983(3)   & 5.714(2) &  5.614(3)\\
30 & 0.9902(1) & 1.0131(8) & 0.990(3)   & 5.734(3) &  5.618(6) \\
\hline
\end{tabular}
\end{table}

\begin{figure}
\includegraphics[width=0.5\textwidth]{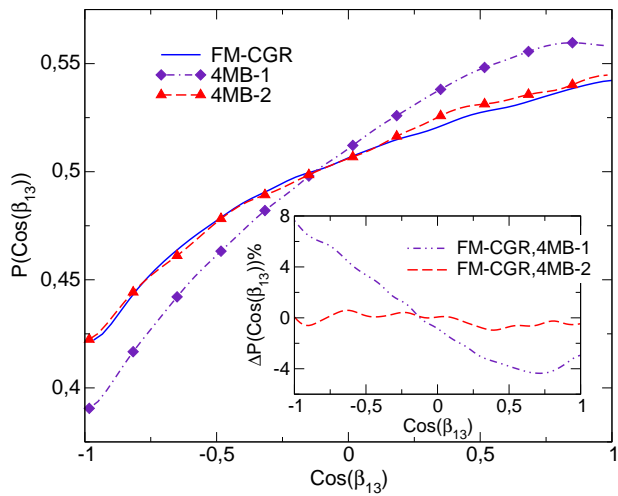}
\caption{Probability distribution of $\cos\beta_{13}$ in the original tetramer
model (4MB-1) and in the modified tetramer model (4MB-2). The distribution for a
CGR of a full-monomer chain with four blobs is also reported for comparison (FM-CGR). The 
inset displays the relative difference of the two models with the 
FM-CGR prediction. } \label{fig:fig1}
\end{figure}
\begin{figure}[htb!]
\begin{center}
\begin{tabular}{ccc}
\epsfig{file=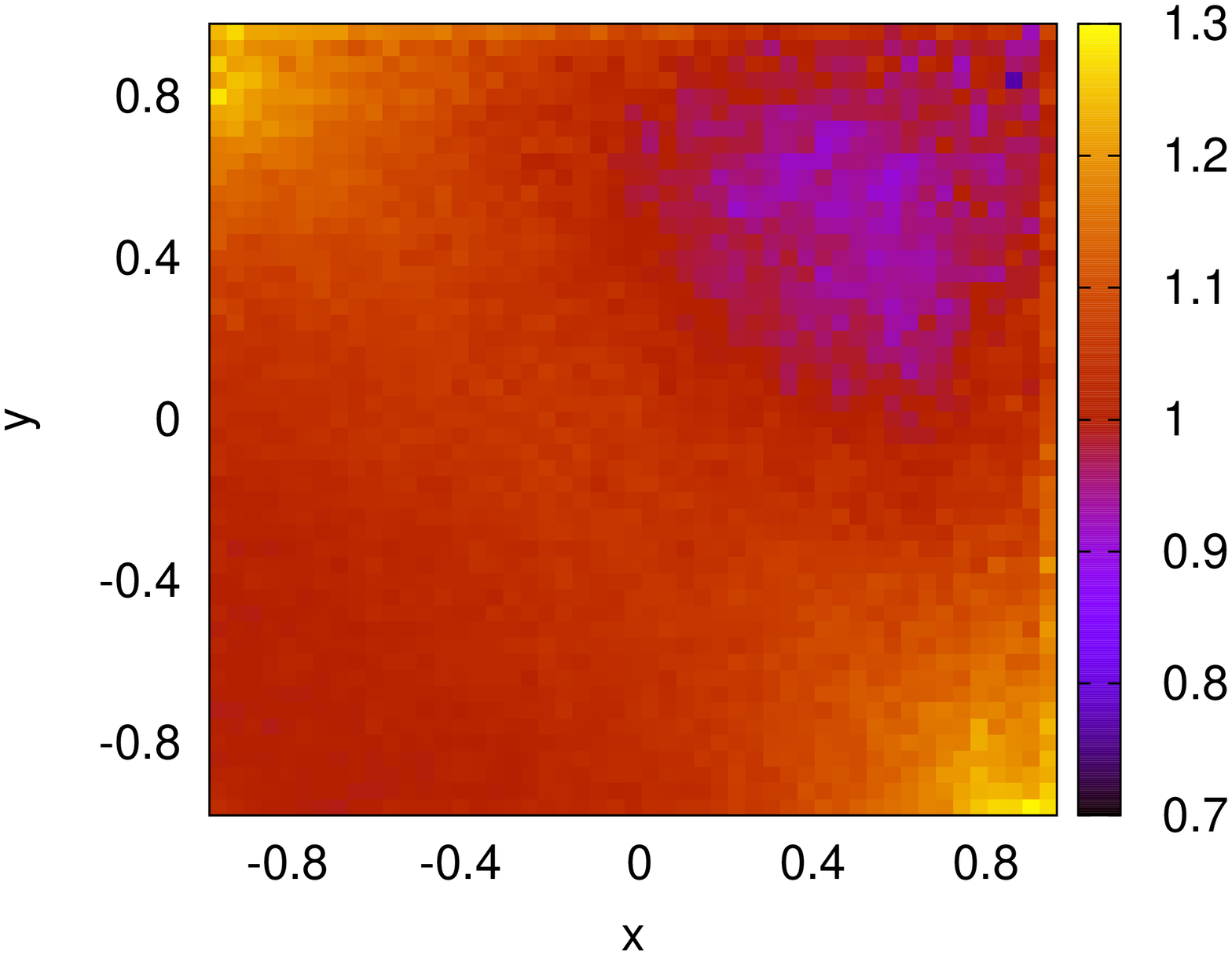,width=0.245\linewidth,angle=-90} & 
\epsfig{file=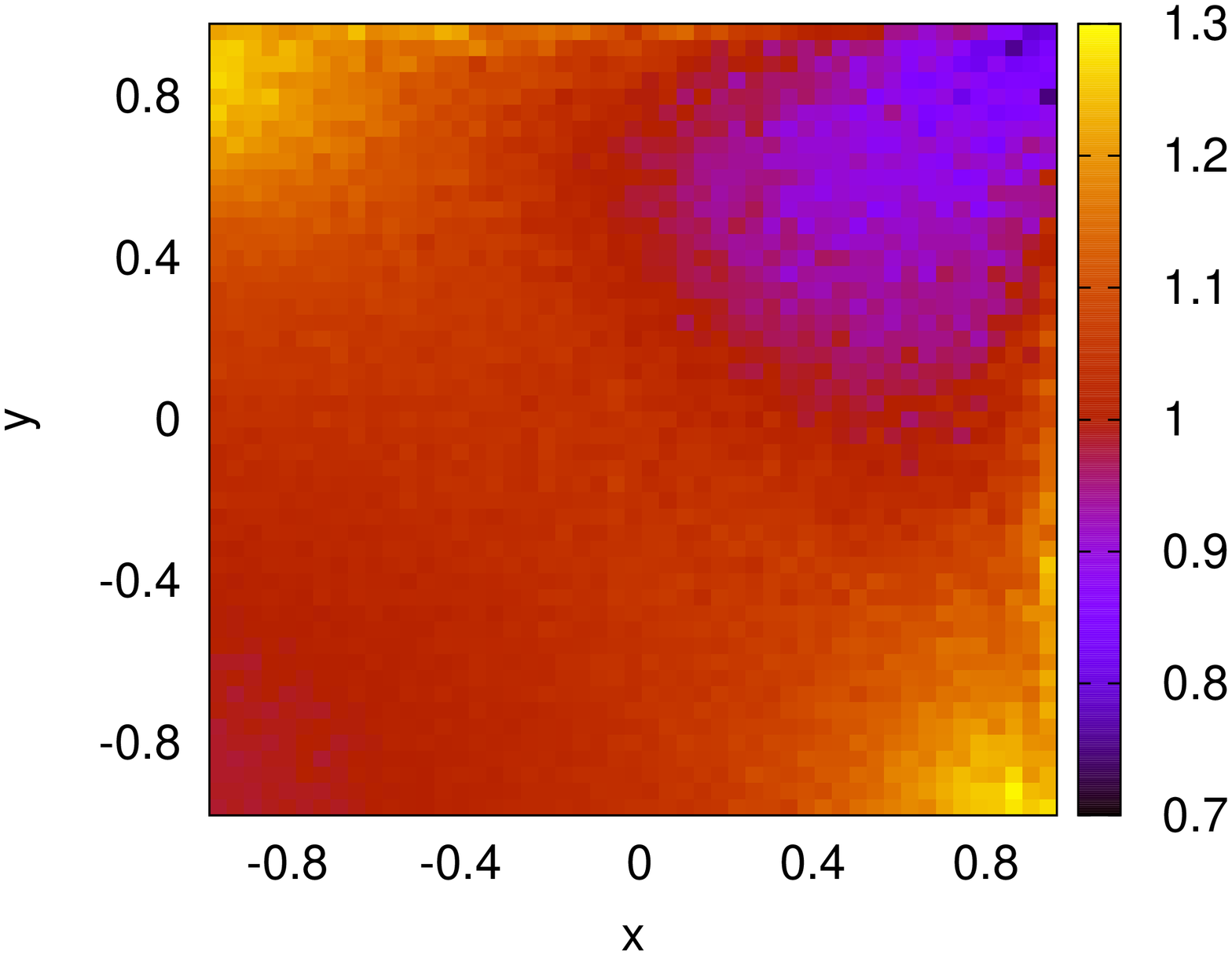,width=0.245\linewidth,angle=-90,clip=} &
\epsfig{file=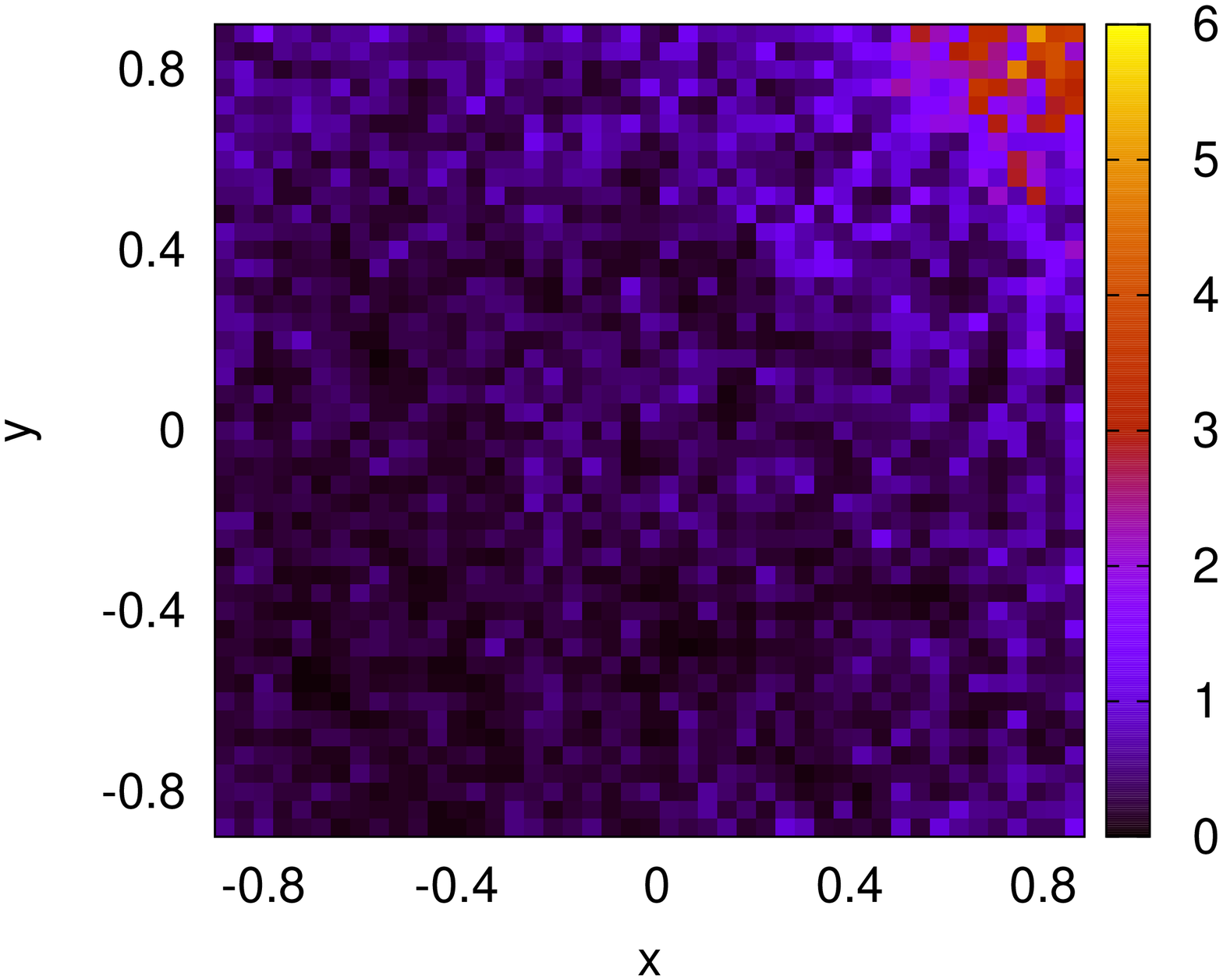,width=0.245\linewidth,angle=-90,clip=}
\end{tabular}
\caption{ 
\1{Cross-correlation function $C_n(x,y)$. Comparison between the 4MB-2 model (left panel) FM-CGR chains (central panel) for $n=4$ at $\Phi=0$. 
In the right panel, we show the relative deviations of the 4MB-2 results from the FM-CGR ones, $\Delta C(x,y)=100|C_{4MB-2}(x,y)/C_{FM}(x,y)-1|$}. }
\label{fig:4body}
\end{center}
\end{figure}

The origin of such inconsistency can be ascribed to a residual inaccuracy 
of the original tetramer model to reproduce four-body correlations. 
In Fig.~\ref{fig:fig1} we show, for $n=4$, the probability 
distribution of
$\cos\beta_{13}=({\bf b_1}\cdot{\bf b_3})/(|{\bf b_1}||{\bf b_3}|)$. 
Data of the 
4MB-1 model are compared with FM-CGR results. Despite the explicit presence of the 
angular potential which reproduces very accurately the torsional angle 
distribution \cite{DAdamoG:2011p2477}, $P(\cos\beta_{13})$ is not well 
reproduced in the 4MB-1 model---more 
elongated configurations are apparently enhanced. 
This inaccuracy, irrelevant for the properties of the 
multiblob model at the tetramer level ($n=4$), accumulates when transferring 
the tetramer potentials to longer chains, producing the observed increase 
of $\hR_{g,b}(n)/\hR_g$ and $A_2$ with $n$. A second, modified 
tetramer model which reproduces the FM-CGR behavior can be obtained by adding a new potential on $\cos\beta_{13}$. 
As for the other potentials, the 
IBI procedure is applied to extract the new optimal potential. At the same time 
the dihedral angle potential is re-optimized to keep the level of 
accuracy of the original tetramer model. 
Explicit expressions for the potentials
are provided in the Appendix. The good 
accuracy of the new model is illustrated in figure \ref{fig:fig1}.
The modified tetramer model can now be safely used as the building block of a 
fully consistent model (4MB-2) which puts in action the ideas of the multiblob 
approach. As illustrated in table \ref{table:Rg}, at zero density the 
agreement with the FM-CGR predictions for the rescaled radius of gyration is 
excellent for all $n$ values investigated. Also the virial coefficient is 
 independent on $n$ and in good agreement with the FM prediction
$A_2 = 5.500(3)$ \cite{Caracciolo:2006p587}. 
This is the first requirement 
of transferability with $n$ at zero density. 
Note that the third virial coefficient appears to be much less sensible to model inaccuracies: we obtain $A_3=\hat{R}_g^6B_3=10.3(2)$ both for 4MB-1 and 4MB-2 for all $n$ values investigated to be compared with the FM prediction\cite{Caracciolo:2006p587}: $A_3=9.80(2)$.

\1{
To check that our parametrization exhausts all four-body terms we compare in Fig. \ref{fig:4body} the cross correlation between the cosine of the two bending angles of the tetramer as obtained by the FM-CGR and by the 4MB-2 model. This quantity is defined as 
\beq
C(x,y)=\frac{\langle\delta(\cos\beta_1-x)\delta(\cos\beta_2-y)\rangle}
{\langle\delta(\cos\beta_1-x)\rangle\langle\delta(\cos\beta_2-y)\rangle}.
\eeq
In the absence of correlation we have $C(x,y)=1$. We observe the presence of a region of depleted correlation near $\beta_1=\beta_2=0$, and two regions of enhanced correlation around $\beta_1=0, \beta_2=\pi$ and $\beta_1=\pi, \beta_2=0$, respectively. The 4MB-2 model reproduces fairly well the results of the FM-CGR (see the right panel), except in a very narrow region around $\beta_1=\beta_2=0$ where statistical noise is also larger, indicating the accuracy of our parametrization.
}

\begin{figure}[htb!]
\begin{center}
\begin{tabular}{ccc}
\epsfig{file=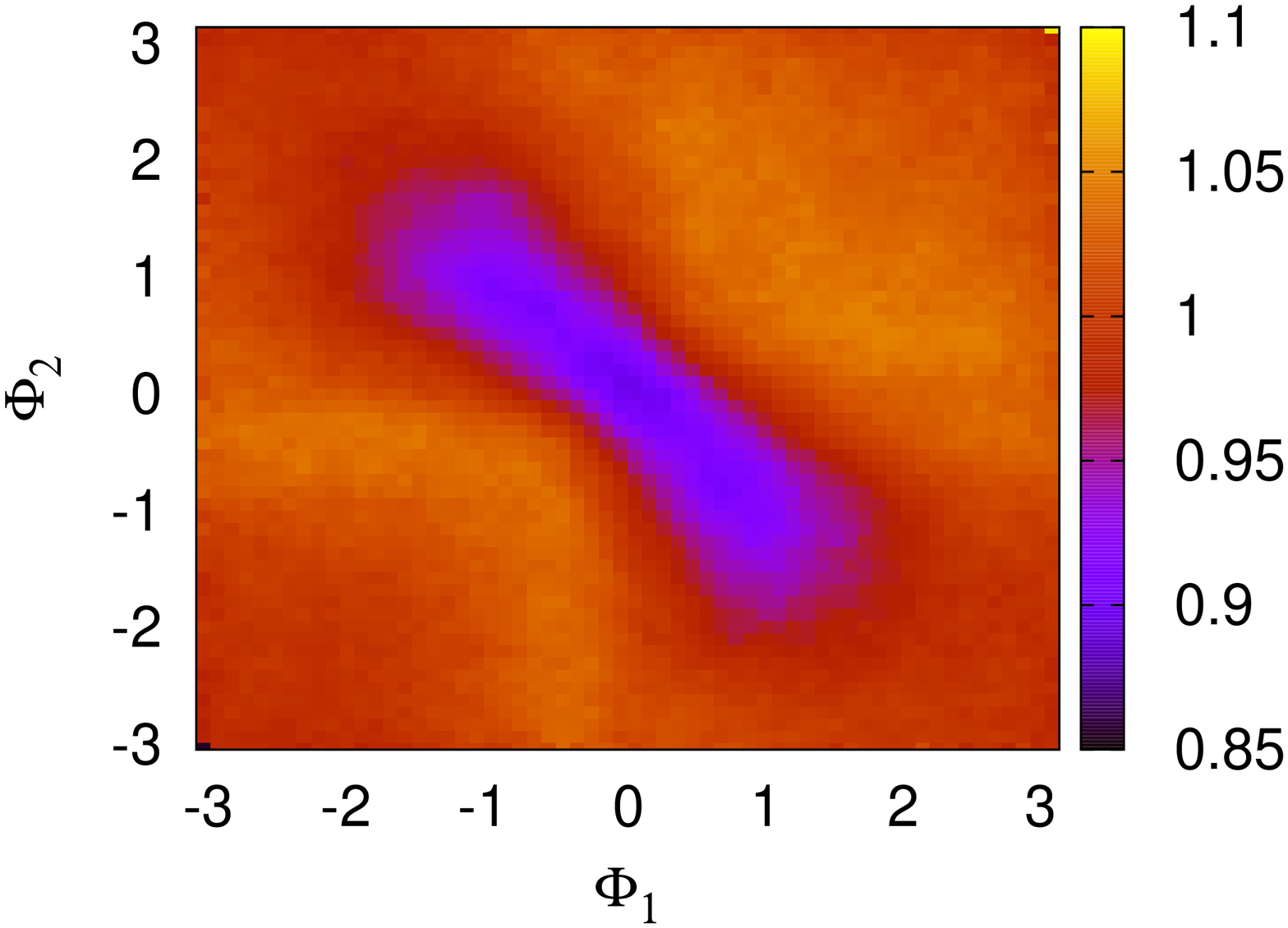,width=0.245\linewidth,angle=-90} & 
\epsfig{file=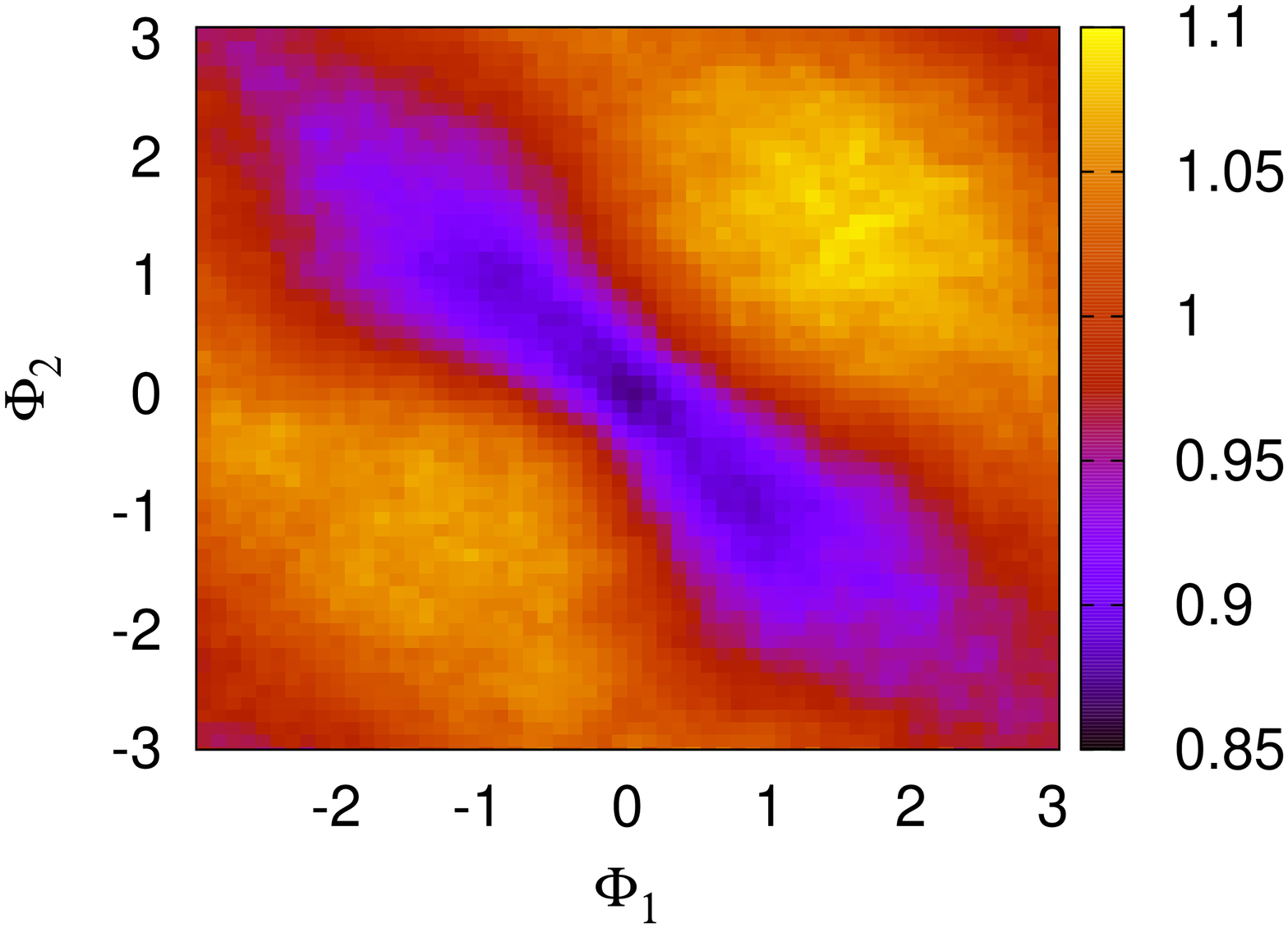,width=0.245\linewidth,angle=-90,clip=} &
\epsfig{file=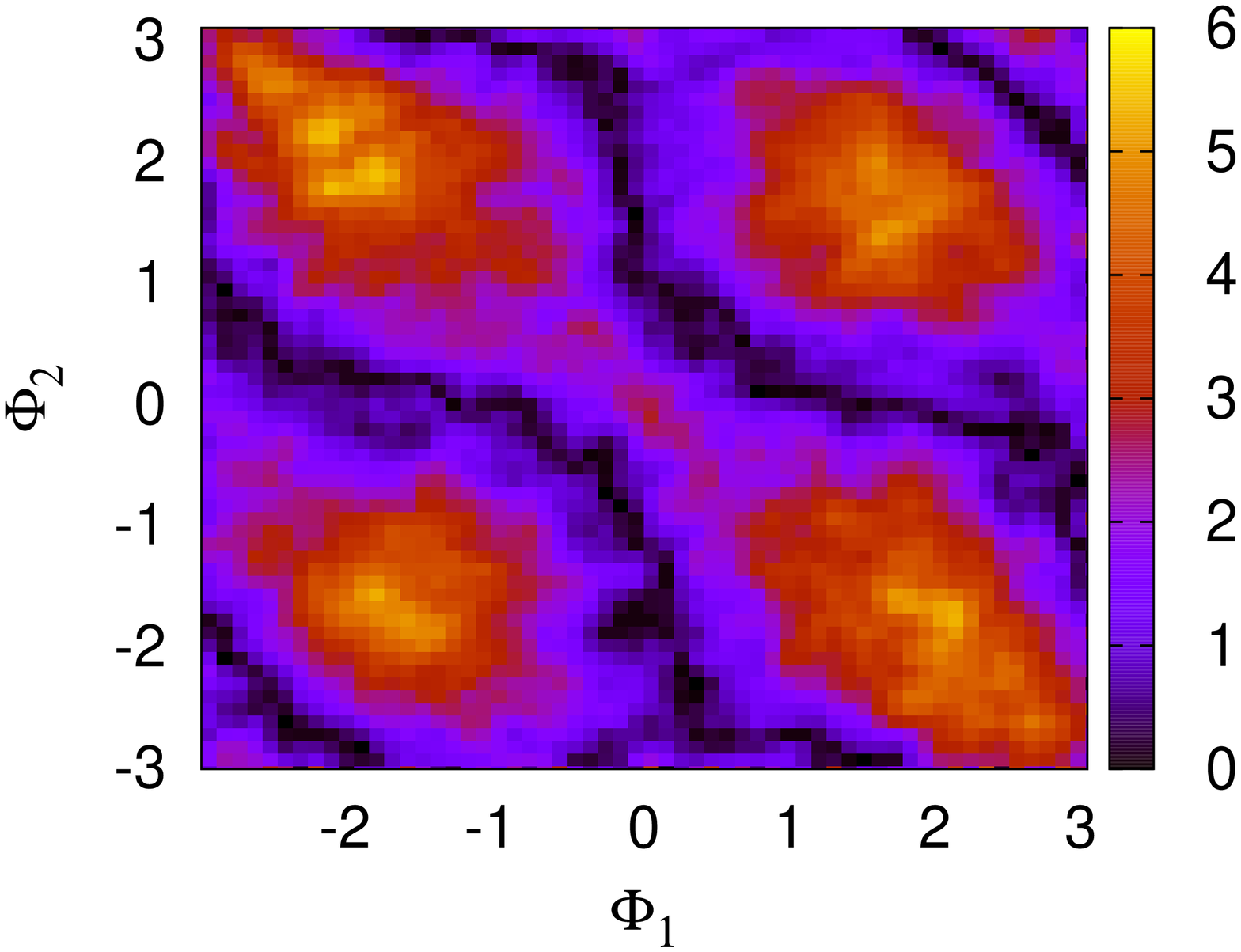,width=0.245\linewidth,angle=-90,clip=} \\
\epsfig{file=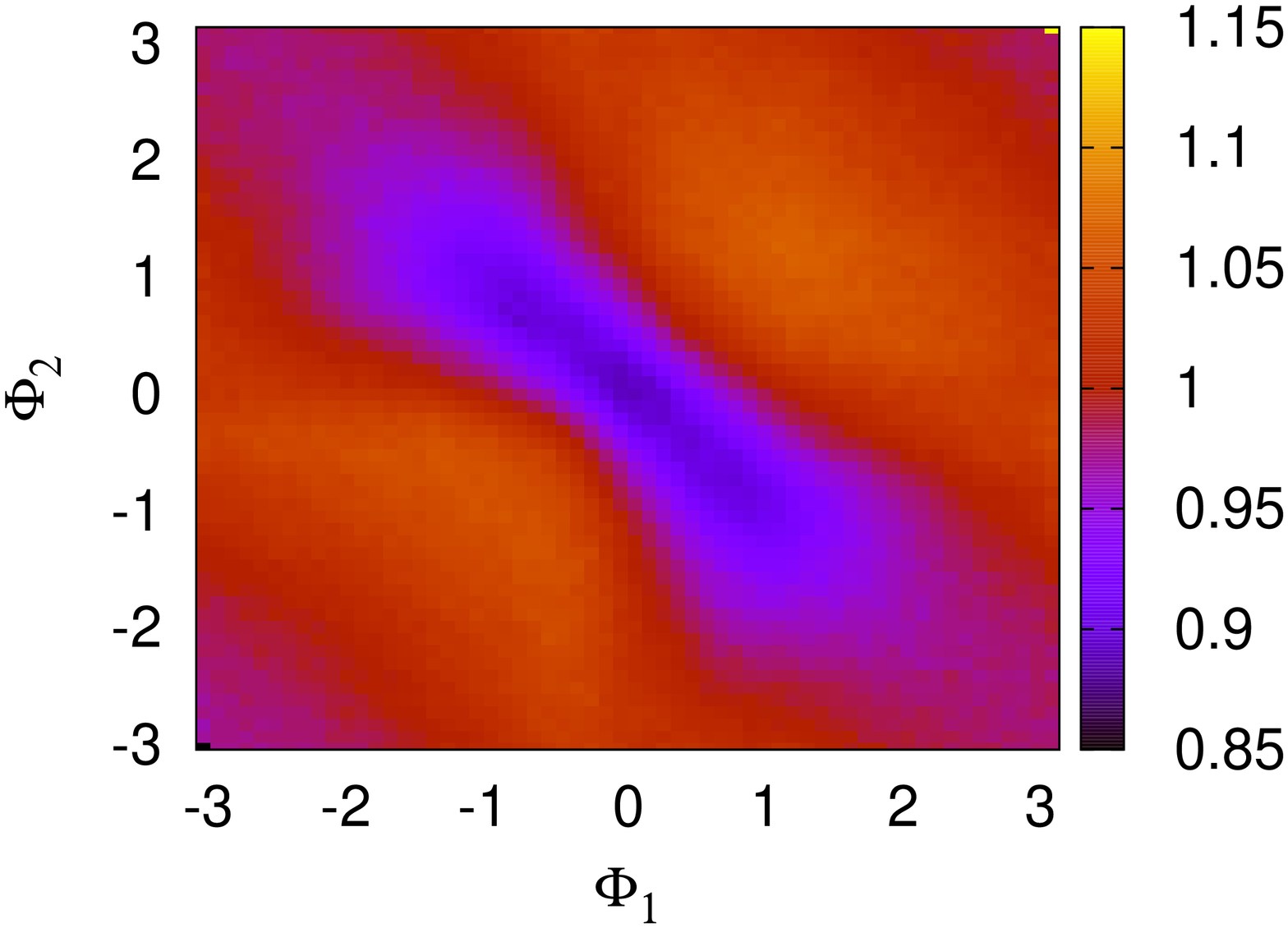,width=0.245\linewidth,angle=-90,clip=} & 
\epsfig{file=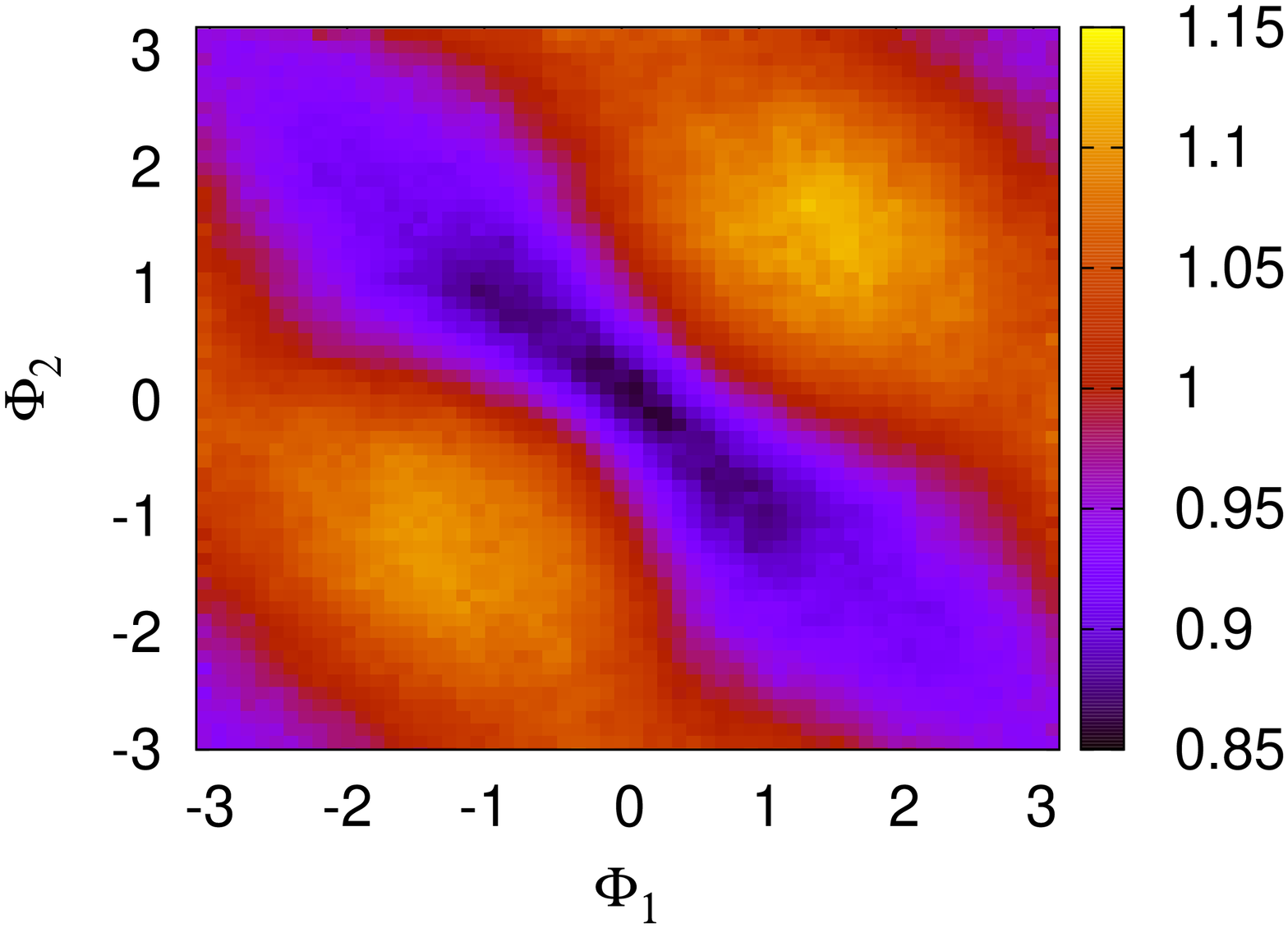,width=0.245\linewidth,angle=-90,clip=} &
\epsfig{file=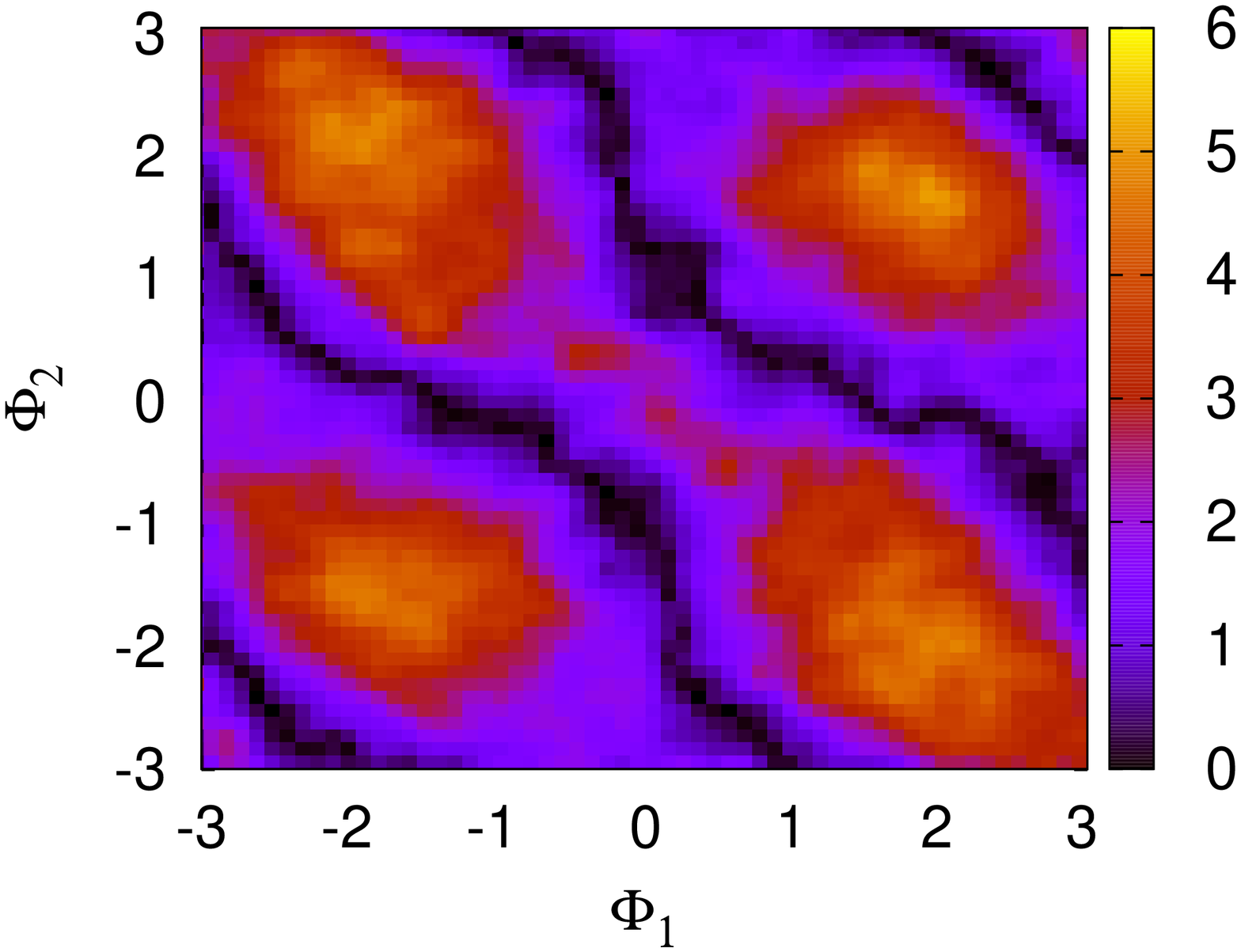,width=0.245\linewidth,angle=-90,clip=} \\
\epsfig{file=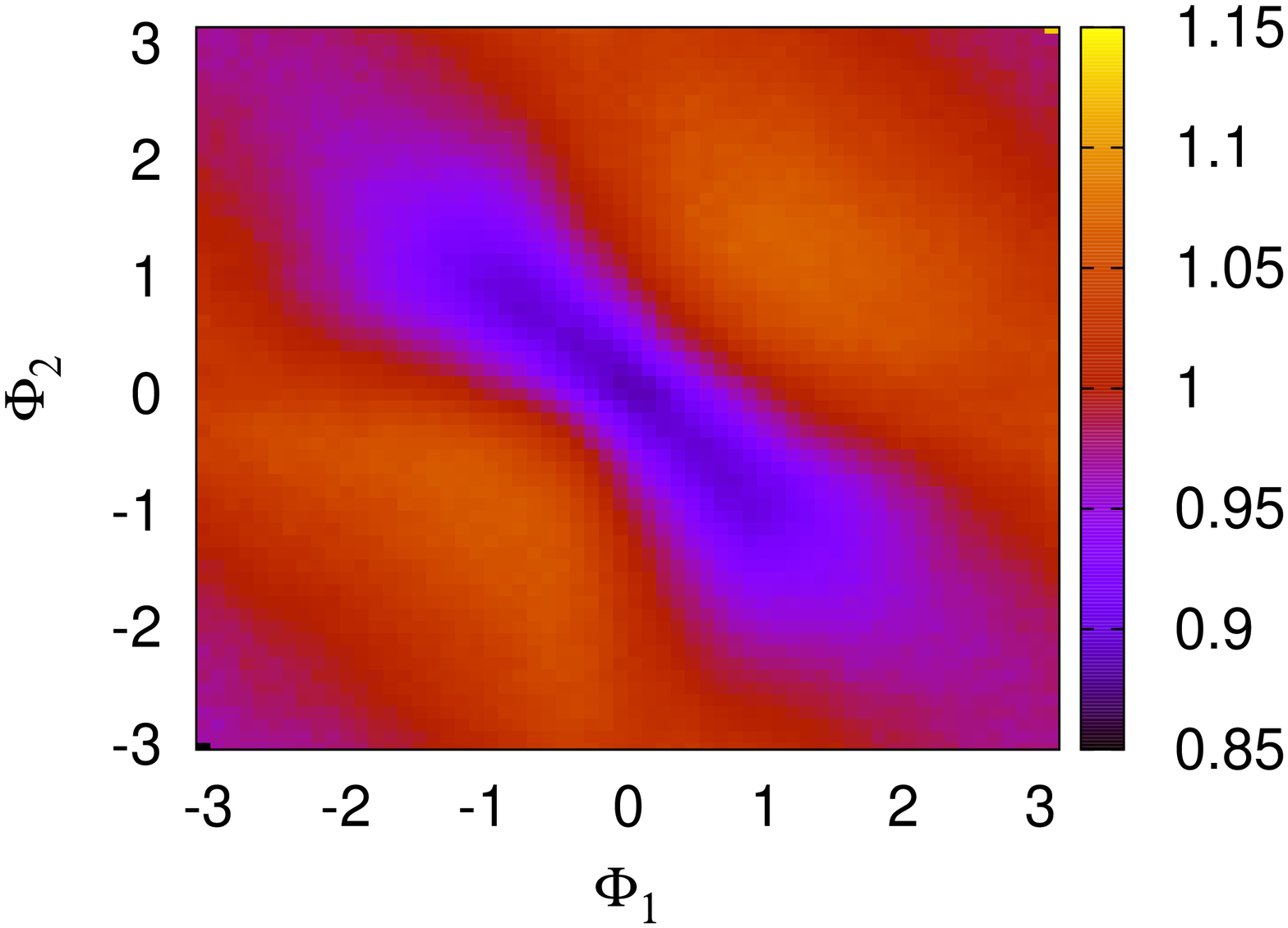,width=0.245\linewidth,angle=-90,clip=} & 
\epsfig{file=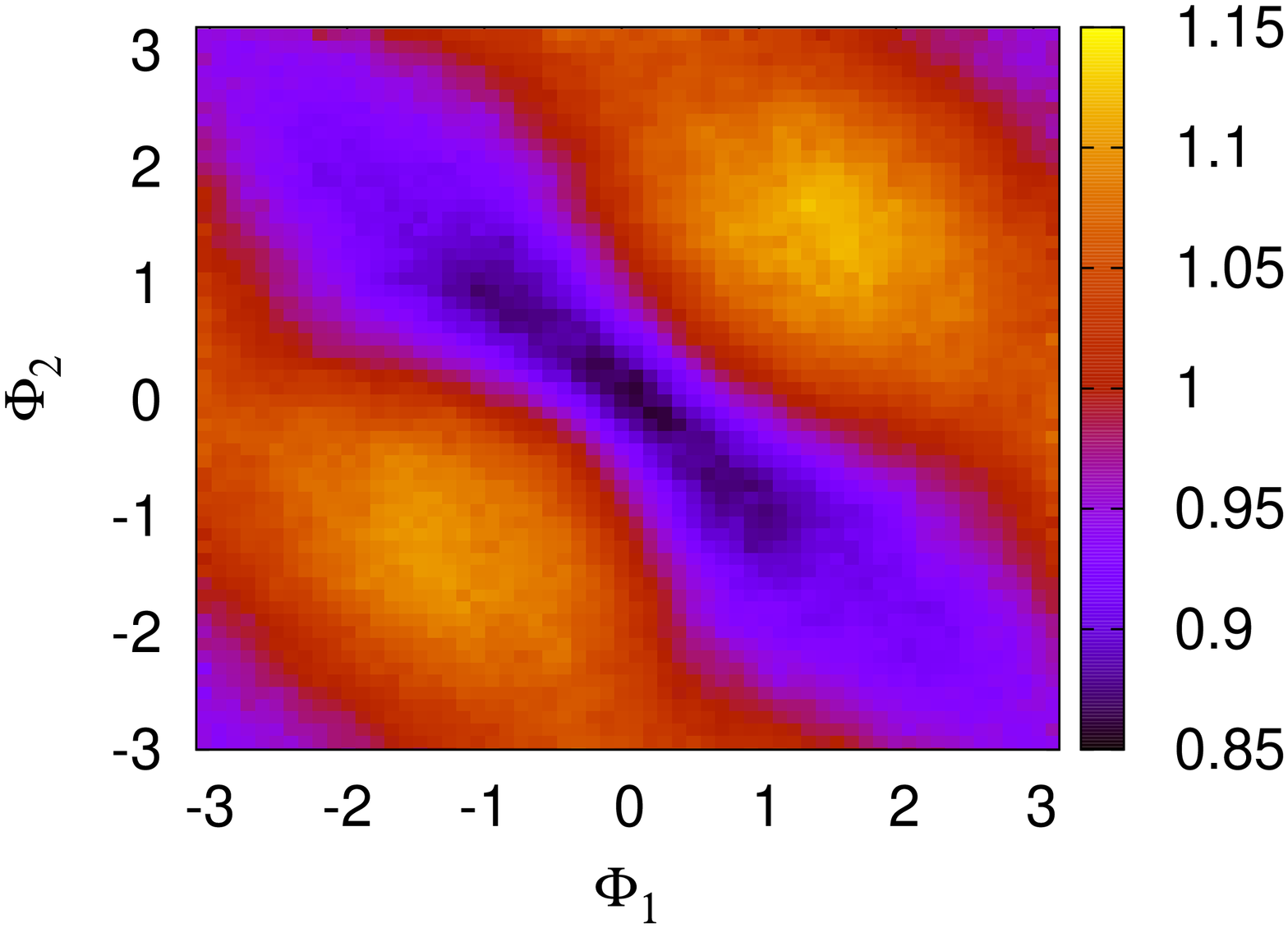,width=0.245\linewidth,angle=-90,clip=} &
\epsfig{file=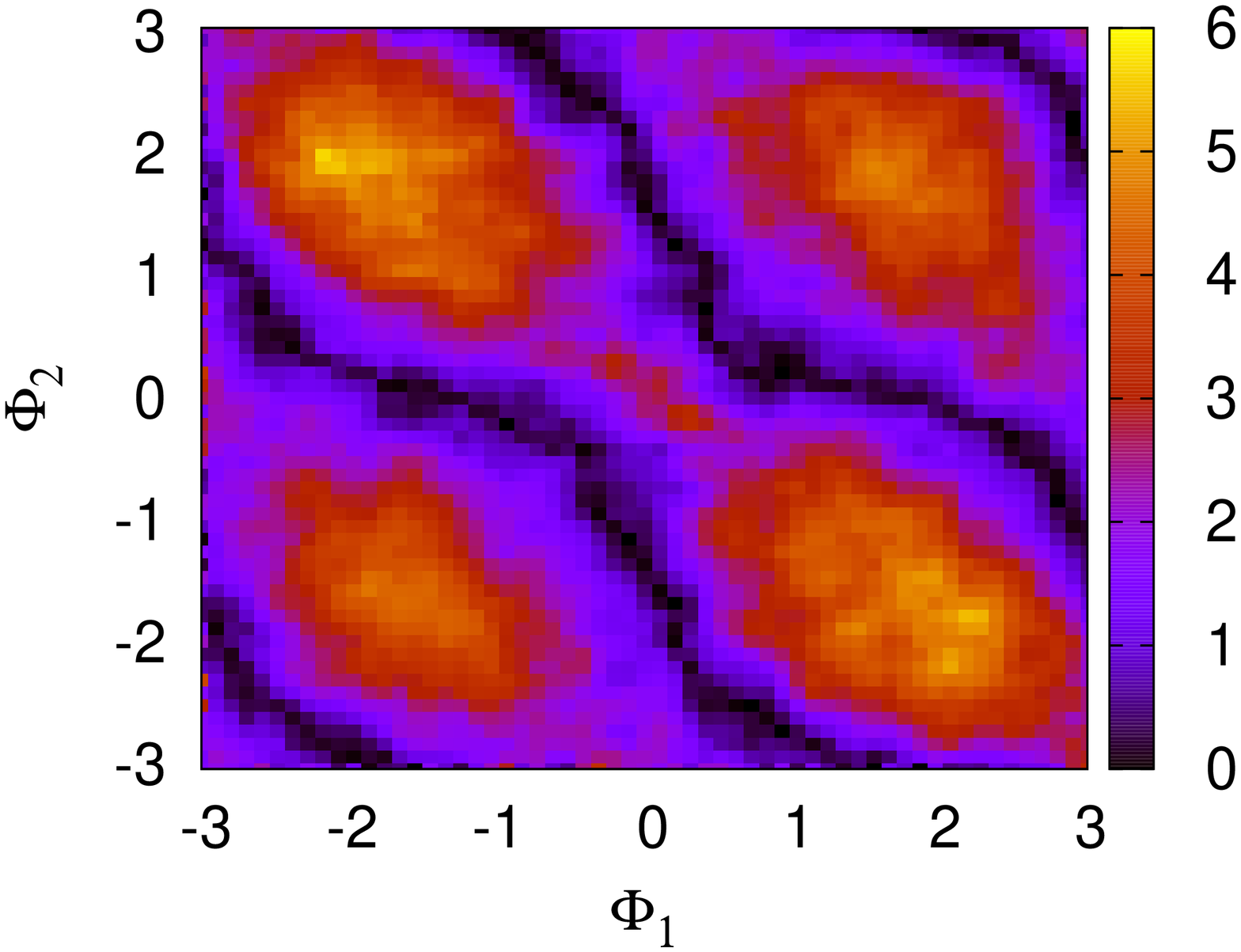,width=0.245\linewidth,angle=-90,clip=} \\
\end{tabular}
\caption{ $\mathcal{H}_n(\Phi_1,\Phi_2)$. Comparison between the 4MB-2 model (left column) and FM-CGR chains (central column) for $n=10$ (upper row), $n=20$ (central row) and $n=30$ (lower row) at $\Phi=0$. On the right column, for each value of $n$, we show the deviations of the 4MB-2 results from the polymer ones, $\Delta(\Phi_1,\Phi_2)=100|\mathcal{H}_{4MB-2}(\Phi_1,\Phi_2)/\mathcal{H}_{FM}(\Phi_1,\Phi_2)-1|$ }
\label{fig:R1}
\end{center}
\end{figure}

To further investigate if of the new model 4MB-2 is able to reproduce FM-CGR correlations, we computed genuine five-body intramolecular correlations for varying number of blobs $n$. Note that two-, three- and four-body correlations at short chemical distance are explicitly built into the model, and five-body correlations are the first missing terms. 
A useful tool to characterize these correlations is the torsional angle correlation function $\mathcal{H}_n(\Phi_1,\Phi_2)$ between two subsequent dihedral angles along the backbone. If $\theta_i$ is the $i$-th torsional angle along the chain,
we adopt the following definition for the function $\mathcal{H}_n(\Phi_1,\Phi_2)$:
\begin{equation}
\mathcal{H}_n(\Phi_1,\Phi_2)= \frac{\sum_{i=1}^{n-4} \langle \delta(\Phi_1-\theta_{i})\delta(\Phi_2-\theta_{i+1})\rangle}{\sum_{i=1}^{n-4}\langle \delta(\Phi_1-\theta_{i})\rangle \langle \delta(\Phi_2-\theta_{i+1})\rangle},
\end{equation}
where $\langle\dots \rangle$ is the average over single-chain configurations, and the subscript $n$ emphasizes the dependence of $\mathcal{H}_n$ on the number $n$ of blobs. It is possible to define similar functions for any pair of torsional angles along the backbone, not necessarily nearest neighbors. Since we expect the correlation between such pairs to decay rather quickly with the chemical distance we limit our analysis to nearest neighbors. Note that with this definition, $\mathcal{H}_n(\Phi_1,\Phi_2)=1$ in the absence of correlation.
In Fig. \ref{fig:R1} we report the function $\mathcal{H}_n$  for $n=10, 20, 30$ in both 4MB-2 (left panels) and FM (central panels) representation and the relative deviations (right panels). The 4MB-2 model predictions of $\mathcal{H}_n$  are quite close to the polymer results, the deviations observed are at most $6\%$  in well localized regions of the \1{$(\Phi_1,\Phi_2)$} plane, and are nearly independent on $n$, confirming the good transferability of the potential, as observed for the other structural properties. This observation supports our implicit assumption that irreducible $n$-body terms in the intramolecular potential for $n\geq 5$ are negligible.

\begin{figure}[htb!]
\begin{center}
\includegraphics[width=1.0\textwidth]{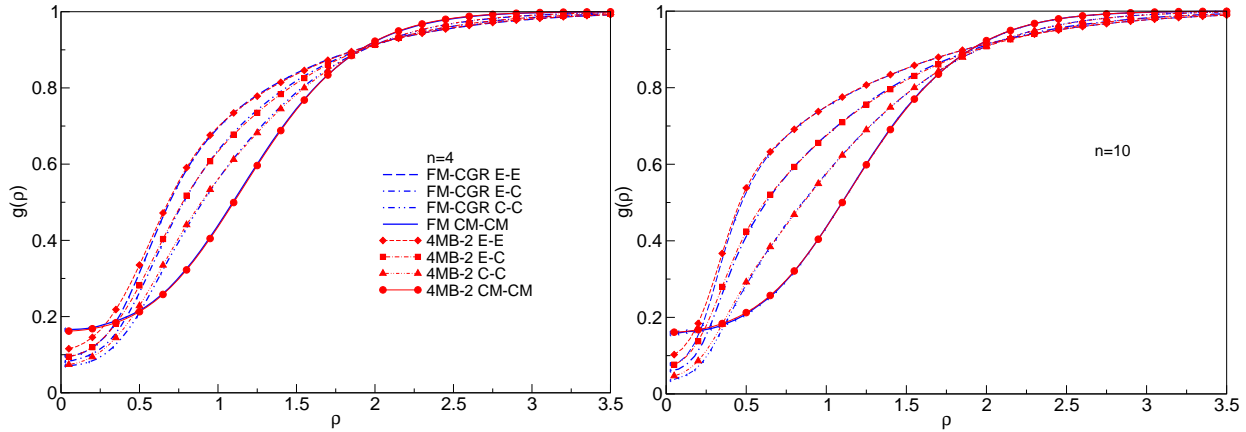}
\caption{Radial distribution functions (RDF) between a pair of isolated chains (zero density) as a function of $\rho=r/\hR_g$. Comparison between FM-CGR and 4MB-2 model predictions for $n=4$ (left panel) and $n=10$ (right panel). RDF between centers of mass (CM-CM), between end monomers (E-E), between central monomers (C-C) and between end and central monomers (E-C) are shown.}
\label{fig:grp}
\end{center}
\end{figure}

It might appear surprising that we spent so much effort in modeling the intramolecular potential while the intermolecular interaction is treated in a very simple way. 
The effective interaction among $N$ chains, of $n$ blobs each, is a general function of $(3 n N - 6)$ variables. According to the multiblob ideas, at any finite reduced chain density $\Phi$, we can adopt the small density expansion and neglect many-chain interaction terms for $n$ large enough. However the effective intermolecular potential between two chains of $n$ blobs remains a function of $6 n  - 6~(n>1)$ variables. 
In our model we adopted several simplifying hypotheses: we replace the intermolecular chain-chain interaction by a sum of pair-wise potentials between blobs of different chains, we assume that all these pair potentials can be adequately approximated by a single central potential, and finally we approximate the potential with a single Gaussian function. It would not be too difficult to avoid the last two assumptions. It is instead crucial, from a practical point of view, to assume pairwise blob-blob interactions. This unique intermolecular blob-blob potential is optimized in such a way to reproduce the polymer center-of-mass radial distribution function (RDF) $g_{CM}$ determined from full monomer simulations \cite{DAdamoG:2011p2477}. This criterion ensures the correct thermodynamics for the CG model since the isothermal compressibility is related to the integral of the RDF by the compressibility equation\cite{HansenMcDonald1987}
\begin{equation}
\left(\frac{\partial \beta \Pi}{\partial c}\right)^{-1} = 1 + \frac{3}{4\pi}\Phi \int (g_{CM}(\rho)-1) d^3{\bf \rho}.
\label{eq:compressibility}
\end{equation}
where $\Pi$ is the osmotic pressure of the solution and $\beta=1/k_BT$. 
An equally good choice as thermodynamics is concerned, would have been to reproduce the RDF between any pair of blobs of different chains since in Eq. (\ref{eq:compressibility}), $g_{CM}$ can be replaced by any other intermolecular RDF \cite{Akkermans:2001p1711}. With our simplifying hypothesis of a unique blob-blob intermolecular interaction we are of course unable to perfectly reproduce RDFs between center-of-mass and between any pair of blobs of the two chains simultaneously. In figure \ref{fig:grp} we show some of those RDFs predicted by the 4MB-2 model and we compare them with the corresponding quantities from FM simulation. We observe perfect agreement for the RDF between the centers of mass but, as expected, some small deviations occur for the other RDFs at distances $\rho\leq 0.5$, a range irrelevant to thermodynamics. These small deviations are the price to pay in order to have a much simplified model, still accurate as thermodynamics is concerned.

\begin{figure}
\includegraphics[width=0.5\textwidth]{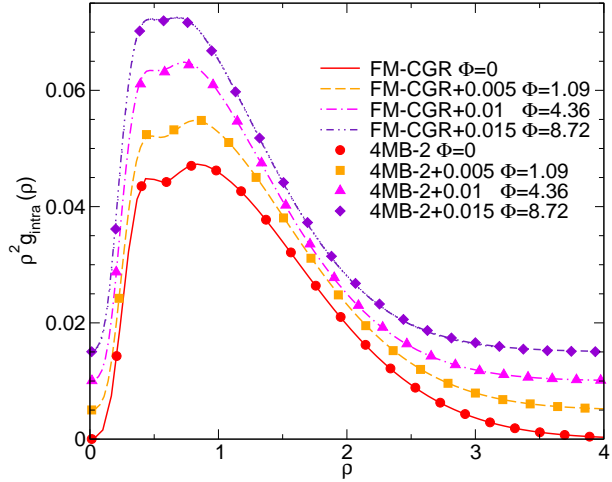}
\caption{Intramolecular radial distribution function $\rho^2 g_{intra}(\rho)$
versus the reduced blob distance $\rho=r/\hat{R}_g$ for model 4MB-2 with $n=20$
and for the FM-CGR chain with 20 blobs. Four values of the reduced density are reported: $\Phi=0, 1.09, 4.36, 8.72$. For sake of clarity, results at different densities are shifted upward according to the legend.} \label{fig:fig2}
\end{figure}

\subsection{Model transferability with density}
The second and more difficult requirement for a transferable MB model is the ability to reproduce FM
predictions at any finite \1{reduced} chain density by increasing the number $n$ of blobs.
To illustrate the ability of our model to match this second
requirement we compare single-chain properties and the Equation of State (EOS)
with FM predictions in the reduced density range $\Phi=c/c^*\lesssim 9$. 
\1{Note that for the present MB model, which is resolution invariant at zero density, 
the definition of the reduced polymer density $\Phi$ is given in terms of an $n$-independent 
characteristic length, the FM radius of gyration $\hat{R}_g$.
This implies that at given values of $\Phi$, MB models with different $n$ are all at the same 
density $c=N/V$ of the underlying polymer model. 
Conversely, when adopting non-resolution invariant models \cite{Pierleoni:2007p193,DAdamoG:2011p2477}, comparing systems with different $n$ at the same value of $\Phi$ implies comparing systems at different values of the absolute polymer density $c$ which is inconsistent.}

\begin{figure}
\includegraphics[width=0.5\textwidth]{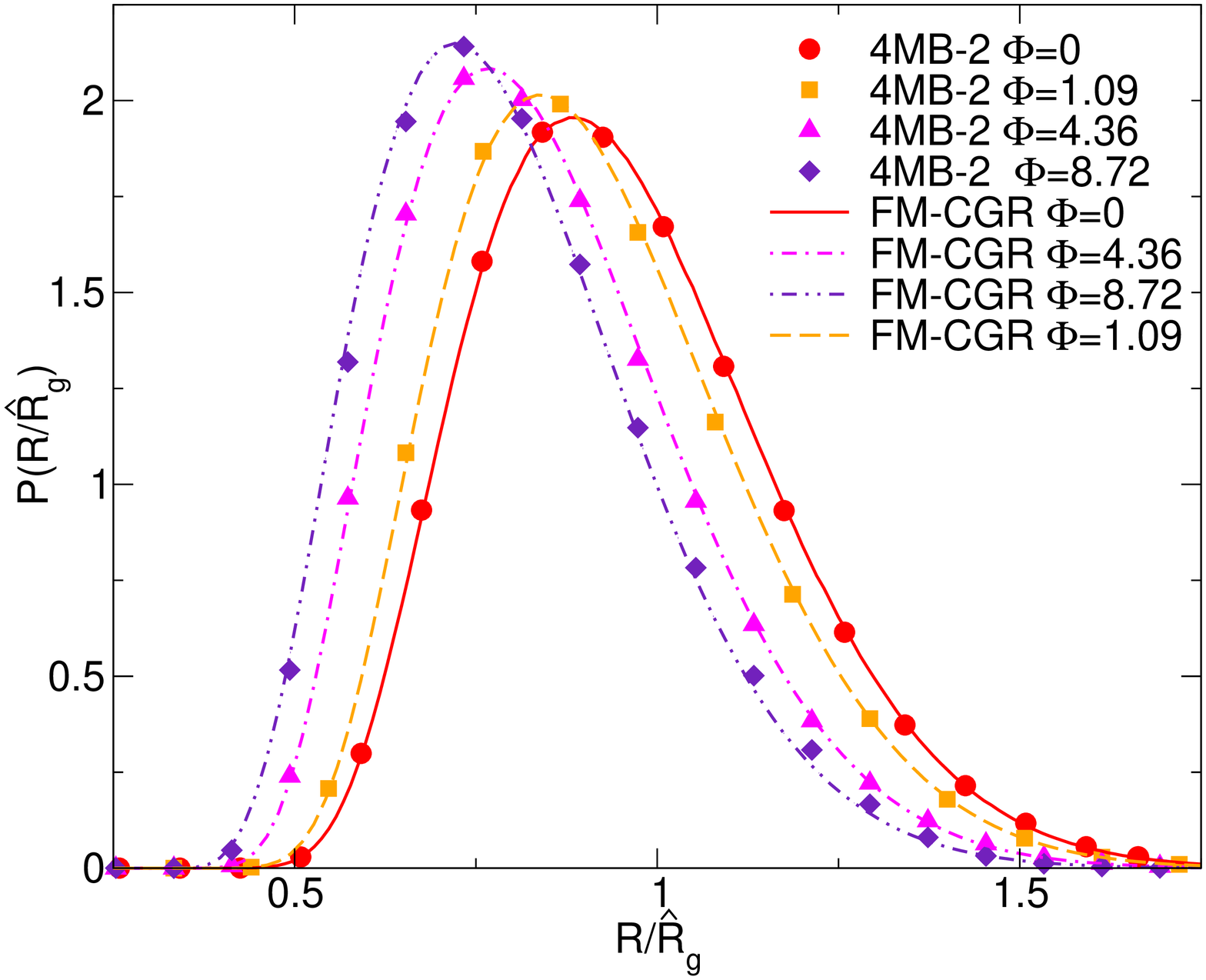}
\caption{Chain radius-of-gyration probability distribution function
$P(R/\hat{R}_g)$ ($R$ is the radius of gyration of a single chain and 
$\hat{R}_g$ its zero-density average) 
at $\Phi=0, 1.09, 4.36, 8.72$ for a chain of $n=20$ blobs.
Comparison between the 4MB-2 model and the FM\1{-CGR} predictions.} \label{fig:fig3}
\end{figure}

In Fig. \ref{fig:fig2} we compare FM\1{-CGR} and 4MB-2 predictions for the 
intramolecular radial distribution function $\rho^2 g_{intra}(\rho)$, which is 
universal in terms of the reduced intra-blob distance $\rho=r/\hR_g$, for a 
chain of $n=20$ blobs and for increasing density. 
The observed agreement is remarkable in particular at the highest density 
investigated, $\Phi=8.72$, deep inside the semidilute regime.
The same kind of agreement is observed for the distribution of the radius of 
gyration $P(R/\hR_g)$ shown in Fig. \ref{fig:fig3}.
The agreement worsens at the highest densities for shorter chains, but it 
remains very good for $n\geq 20$.
We would like to emphasize the nontrivial nature of the observed agreement. 
Indeed, $\rho^2 g_{intra}(\rho)$ is a weighted average of the length 
distribution over all pairs of blobs, so the agreement for this quantity 
demonstrates the accuracy of the 4MB-2 model 
to reproduce pair distances well beyond 
four neighbors. $P(R/\hR_g)$ includes information on pair distances but also 
on all possible cross-terms between different pairs, so the observed 
agreement for this quantity is even less obvious.

Finally, we have computed the EOS, using the standard molecular virial route 
to the pressure for several values of $n$ \cite{Akkermans204}. 
In Table \ref{table:eos}
we report the compressibility factor $Z=\beta \Pi/c$, where  
$\Pi$ is the osmotic pressure of the solution. 
\begin{table}
\caption{Compressibility factor $Z$ for the 4MB-2 model for several values
of $n$ and for polymers in the scaling limit
\protect\cite{Pelissetto:2008p1683}, at various reduced densities 
$\Phi$ from the dilute to the semidilute regime.} 
\label{table:eos}
\begin{tabular}{ccccc} \hline
$\Phi$  &  $n=10$ &  $n=20$ & $n=30$ & FM \\
\hline
0.54  & 1.875(1)    & 1.874(2)   & 1.870(2)   & 1.85(1) \\
1.09  & 2.9760(4)  & 2.9784(6) & 2.9819(1) & 2.96(1) \\
2.18  & 5.6004(7)  & 5.708(1)   & 5.729(3)   & 5.63(1) \\
4.36  & 11.4541(5)& 12.058(1) & 12.269(3) & 12.3(1) \\
6.54  & 17.532(2)  & 18.931(4) & 19.515(5) & 20.0(1) \\
8.72  & 23.751(1)  &  25.592(4) & 26.989(8) & 28.7(1) \\
\hline
\end{tabular}
\end{table}
Up to density $\Phi\simeq 2$, results for all values of $n$ studied  
differ by less than 1\% from 
the asymptotic polymer value. When density increases longer chains are systematically 
more accurate. At $\Phi=4.36$ the relative deviation in $Z$ is $\sim 5\%$ 
for $n=10$, but $\lesssim 1\%$ for both $n=20$ and $n=30$. At $\Phi=8.72$, 
the deviation is $\sim 15\%$ for $n=10$, and decreases to $\simeq 10\%$ for 
$n=20$ and $\simeq 6\%$ for $n=30$. 
\1{This is expected from the multiblob heuristic argument since from Eq. (\ref{equation:k2}) the reduced blob density,
\beq
\Phi_b(n)=\frac{c_b}{c_b^*}=\left(1.03-\frac{0.04}{n}\right)^3 n^{1-3\nu}~\Phi,
\label{eq:phib}
\eeq
decreases, at fixed $\Phi$, for increasing $n$ as long as $\nu>1/3$. At $\Phi=8.72$, $\Phi_b(10)\simeq 1.6,  \Phi_b(20)\simeq 1.0$ and $\Phi_b(30)\simeq 0.7$ which explains why increasingly large $n$-values are necessary in order to neglect three- (or more)-blobs contacts as in the present model.}

\1{It is of some interest to investigate the relation between the blob radius of gyration of the coarse grained representation $\hr_g(n)$ and the density-dependent correlation length $\xi$ characteristic of semidilute polymer solutions \cite{Doi}, also called correlation blob size. The latter quantity can be inferred from the single chain structure factor, $S_{intra}(q)=L^{-2}\langle\sum_{ij}e^{-i\qvec\cdot\rvec_{ij}}\rangle$, as the characteristic length at which a crossover from the ideal behavior $(\sim q^{-2}$) to the excluded-volume behavior ($\sim q^{-1/\nu}$) is observed \cite{Farnoux78,Paul91,Muller00}. In figure \ref{fig:kratky} we show at $\Phi=8.72$ the Kratky plot, $(qR_g)^2S_{intra}(q)$ against $qR_g(\Phi)$, for the full monomer Domb-Joyce model (FM), the FM-CGR with $n=30$ and the 4MB-2 model with the same number of blobs. In the FM-CGR case $S_{intra}(q)$ is defined as above with monomer positions $\rvec_i$ replaced by the blob center-of-mass positions $\bs_i$.
\begin{figure}
\includegraphics[width=0.6\textwidth]{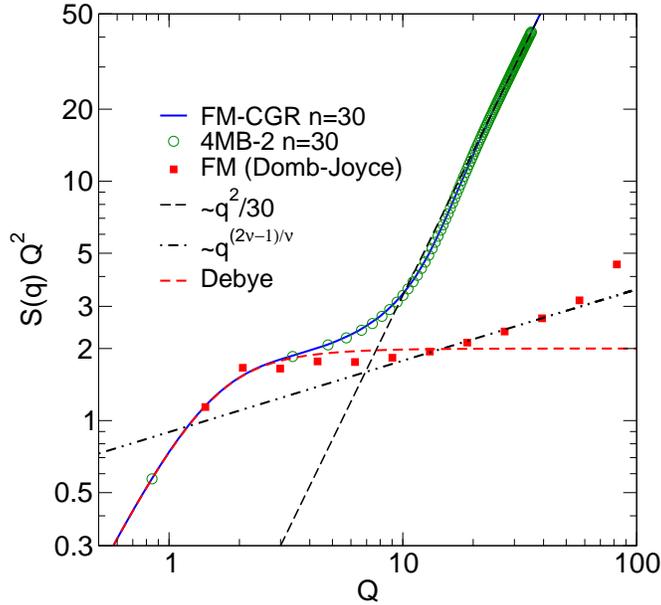}
\caption{\1{ (color online) Kratky plot at $\Phi=8.72$ as a function of $Q=qR_g(\Phi)$. Full monomer Domb-Joyce model (FM, red squares), the same chain model mapped onto a 30-blobs chain (FM-CGR, blue line), the model 4MB-2 with $n=30$ blobs (open green circles), the Debye function (dotted line). Note that $Q=qR_{g,b}(\Phi)$ has been used for the FM-CGR and 4MB-2 models. For $n=30$ the difference between between $R_g$ and $R_{g,b}$ is already small and will become even smaller for larger values of $n$.}} \label{fig:kratky}
\end{figure}
The correlation blob size $\xi$ can be readily determined from the FM data to be 
$\xi/R_g(8.72) \sim 2\pi/10\simeq 0.63$, which corresponds to $\xi/\hR_g\simeq 0.53$ since $R_g(8.72)/\hR_g=0.8423$ 
at this reduced density. 
On the other hand from Eq. (\ref{equation:k2}) we obtain $\hr_g(n)/\hR_g=k/n^{\nu} = 0.27, 0.18$, and $0.14$ for $n=10, 20$, 
and $30$, respectively, which are considerably smaller than $\xi$. 
This result is consistent with the multiblob argument: from Eq. (\ref{eq:phib}),
$\Phi_b\propto n^{(1-3\nu)} \Phi=(\hR_g/r_g)^{(1-3\nu)/\nu}\Phi$. 
On the other hand in the semidilute regime \cite{Doi} $\xi/\hR_g\propto\Phi^{\nu/(1-3\nu)}$. Substituting the former relation in the latter we get: $\Phi_b\propto(r_g/\xi)^{(3\nu-1)/\nu}$.
Therefore, as far as $\nu> 1/3$, $\Phi_b\ll 1$ implies $r_g\ll\xi$ in agreement with our findings.
 
In Fig. \ref{fig:kratky} we note that the CGR at such resolution reproduces FM data up to $Q\simeq 2$. If one wishes to have access to more local properties he should take a CGR with more blobs. 
In order to follow the crossover from ideal to excluded volume behavior in the CGR a quite larger number of blobs should be considered. Note that the large-$q$ behavior of the CGR ($\sim q^2$) is universal since our blobs contain a sufficiently large number of monomers to be representative of the scaling limit. Therefore this large-$q$ behavior is inherent to the CGR and not to the particular MB model we considered, namely the 4MB-2 model. Finally, note the that perfect agreement between the 4MB-2 prediction and the FM-CGR data for the single chain structure factor is fully compatible with previous results on $g_{intra}(r)$ (see Fig. \ref{fig:fig2}).
 }

\section{Conclusions}\label{sec:conclu}
In conclusion, we have developed a fully consistent, scale preserving, 
multiblob model for linear polymers in good solvent. The model is built by
transferring to smaller length scales a tetramer model parametrized to
reproduce a number of scalar correlations of the FM chain at zero density. We have shown that this model is fully consistent when varying the number of blobs at zero density, and more relevant, it is able to reproduce the universal EOS for athermal semidilute polymer solutions at high chain concentrations if a sufficiently large numbers of blobs per chain is chosen \1{in such a way to always remain in the dilute blob regime}.
In particular, the present multiblob model with only 30 blobs provides the compressibility factor at $\Phi\sim 9$ with an 
accuracy of $\sim 5\%$, a level 
of accuracy which would require the use of the order of thousand monomers even 
with the most efficient lattice model, namely the Domb-Joyce model tuned in 
such a way to cancel the leading-order corrections to scaling 
\cite{Pelissetto:2008p1683}. 
The present CG strategy may be extended to more complex systems and thus opens the way to a quantitative study of semidilute polymer
solutions in situations where local structure is important like, 
for instance, colloid-polymer mixtures, polymer-decorated colloidal systems and polymer brushes \1{at relatively low grafting density}. It would also be extremely interesting to develop an analogous multiblob strategy to describe the cross-over between the good solvent and the $\theta$ solvent conditions and finally to extend this approach to diblock copolymer solutions. Work in these directions is in progress.

As a final remark it is interesting to observe that our model represents, to a very good
approximation, the first example of a truly "fixed-point" model in the 
renormalization-group language, 
the elusive model which reproduces the scaling behavior at any level of 
coarse-graining. Previous work focused on defining models in which the 
leading scaling corrections were zero within errors. Here, instead, 
we obtain a model in which {\it all} scaling corrections are apparently small,
and which is thus able to predict the scaling behavior for any $n\ge 4$. 
Of course,
there is a price to pay: for any $n$, we do not have access to all possible
observables, but only to the large-scale properties that can be 
modelled by the chosen CG representation.

\section{Acknowledgement}
We thank J.-P. Hansen for inspiring discussions.
CP is supported by IIT under the SEED project grant n 259 SIMBEDD.

\section{Appendix: Explicit expressions of the tetramer potentials} \label{SectionPot}

In this appendix we provide full details on the potentials of the two models used in the present work, namely 4MB-1 and 4MB-2. Distance in the central potentials is $\rho=r/\hat{R}_g$. 

The bonding potentials $\beta V_{12}(\rho;4) = \beta V_{34}(\rho;4)$ and 
$\beta V_{23}(\rho;4)$ have been parametrized for $\rho\leq 3$ as 
\begin{equation}
\beta V_{ij}(\rho;4)=
  (V_0-a_0-a_1)\exp (-a_2\rho^2)+a_1\exp(-a_3\rho^2)+a_4\rho^2+a_5\rho^4+a_0 ,
\end{equation}
where $V_0$ is the value of the potential at the origin, $a_0$ is
a constant which has been fixed so that the potential vanishes at the 
minimum, and we only consider the range $\rho \le 3$.

\begin{figure}[h]
\includegraphics[width=1.0\textwidth]{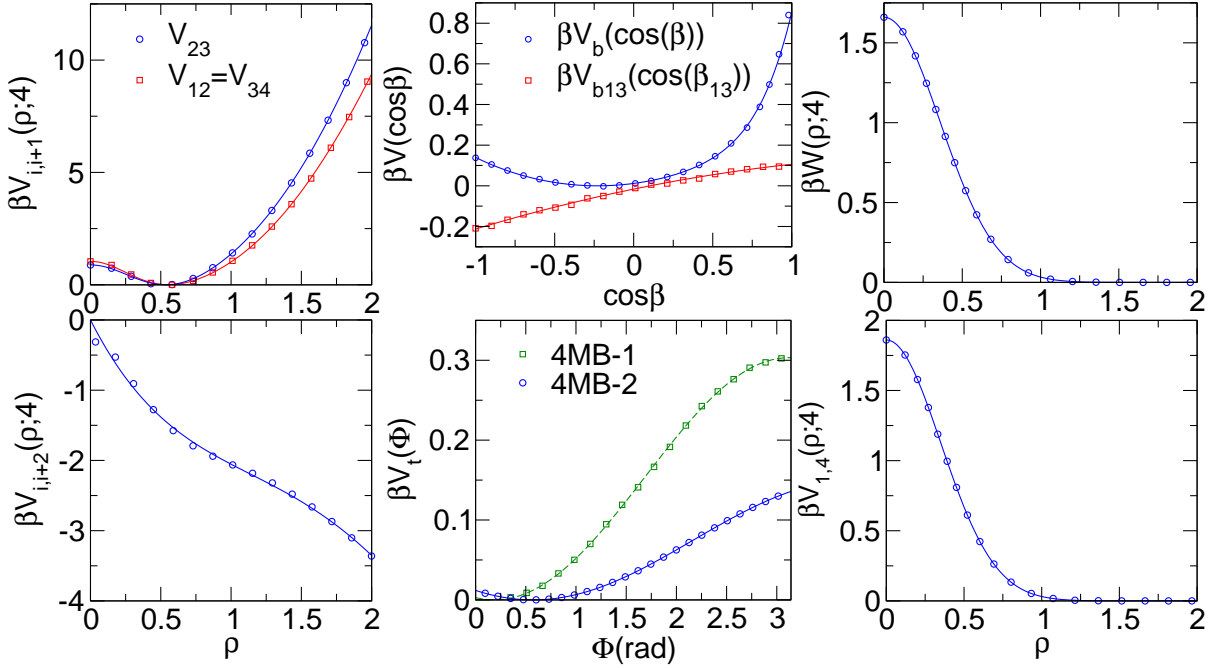}
\caption{Potentials for the tetramer 4MB-2 and 4MB-1. In the left column we show
the pair potentials between first-neighbors along the chain (top)
and between next-to-nearest neighbors (bottom), as a function of $\rho = r/\hat{R}_g$. 
In the central column we report the bending-angle potential, the potential acting on the $cos(\beta_{13}) $ as a function of 
$\cos \beta$ (top), and the torsion-angle potential as a function of $\Phi$ (in radians) for both the models 4MB-1,4MB-2 (bottom). In the right column   we report the intermolecular  pair potential (top) and the potential between the first and the last atom of the tetramer (bottom)  as a function of $\rho = r/\hat{R}_g$. 
The symbols represent the numerical results obtained through IBI procedure and the lines are the corresponding interpolations.}
\label{fig:pot}
\end{figure}

\begin{table}[h]
\caption{Numerical coefficients for the tetramer potentials ($n=4$) for 4MB-1 and 4MB-2 models}
\label{table-coeff-pot}
\begin{center}
\begin{tabular}{ccccccccc}
\hline\hline
 Model & Potential &$V_0$ & $a_0$ & $a_1$ & $ a_2$   & $ a_3$  & $ a_4$   & $ a_5$  \\
\hline
4MB-1,4MB-2 &$V_{23}$ & 0.8808 & $-$6.8935&   1.8693 & 0.4287 & 5.3453 & 4.4274&$-$0.0202\\
4MB-1,4MB-2 &$V_{12}$ & 1.0392 & $-$3.9339&   1.54603& 0.7602 & 6.8442 & 3.3646&$-$0.0194\\
4MB-1,4MB-2 &$V_{13}$ &        &          &$-$3.9518 & 2.9224 &$-$1.1555& 0.1321  \\
4MB-1,4MB-2 &$V_b$    &        &          &   0.2238&$-$0.2284& 0.0769& 0.0600  \\
4MB-1  &$V_t$    &        &    0.1425&   1.0698 &  4.5475& 0.0391& 0.14345 \\
4MB-2 &$V_t$    &        &    $-$0.0797&  0.970 &  0.990& -0.0251& 0.0785 \\
4MB-2 &$V_{b_{13}}$    &        &          &   0.1593&$-$0.0371& &$-$0.0158   \\
\hline

\end{tabular}
\end{center}
\end{table}

The potentials $V_{13}(\rho;4) = V_{24}(\rho;4)$ between next-to-nearest 
neighbors has been parametrized as ($\rho \le 3$)
\begin{equation}
 \beta V_{13}(\rho;4)=\beta V_{24}(\rho;4) = 
   a_1\rho+a_2\rho^2+a_3\rho^3+a_4\rho^4.
\end{equation}
The bending potential is parametrized as 
\begin{equation}
 \beta V_b(\cos\beta)=
   a_1(\cos\beta-a_2)^2+a_3(\cos\beta-a_2)^6+a_4(\cos\beta-a_2)^7
\end{equation}
(the minimum of the potential corresponds to $\cos\beta = a_2$),
while the torsion potential is given by
\begin{equation}
 \beta V_t(\Phi)=a_0(1+a_3\Phi)\sin(a_1 \Phi +a_2) + a_4.
\end{equation}
The potential acting on the $\cos \beta_{13}$ has been parametrized as:
\begin{equation}
\beta V_{b_{13}}(\cos \beta_{13})=a_1\cos \beta_{13}+a_2\cos^2 \beta_{13}+a_4 .
\end{equation}
Numerical values for the coefficients for $n=4$ are reported in Table~\ref{table-coeff-pot} and 
the potential are displayed in figure \ref{fig:pot}. 

Finally, 
 \begin{equation}
 \beta V_{14}(\rho;4)=1.86\exp (-4.08425\rho^2), 
\qquad 
 \beta W(\rho;4) = 1.66\exp (-3.9\rho^2).
 \end{equation}


\newpage

\begin{thebibliography}{10}

\bibitem{Voth2009}
G. Voth, ed., \textit{Coarse-Graining of Condensed Phases and Biomolecular Systems} (CRC Press, Boca Raton, 2009).

\bibitem{PCCP2009}
Themed issue, Phys. Chem. Chem. Phys. \textbf{11}, 1853 (2009).

\bibitem{SM2009}
Themed issue, Soft Matter \textbf{5}, 4341 (2009).

\bibitem{FarDisc2010}
Themed issue, Faraday Discussion \textbf{144}, 1 (2010).

\bibitem{Carbone:2008p2254}
P. Carbone, H. Varzaneh,X. Chen, F. J.  M{\"u}ller-Plathe, J. Chem. Phys. \textbf{128}, 064904  (2008).
  
\bibitem{Fritz:2009p1721}
D. Fritz, V. A. Harmandaris, K. Kremer, and N. F. A. van der Vegt, Macromolecules \textbf{42}, 7579 (2009).  
\bibitem{Peter:2009p1734}
C. Peter and K. Kremer, Soft Matter \textbf{5}, 4357 (2009).

\1{
\bibitem{Tschop98}
W. Tsch\"op, K. Kremer, J. Batoulis, T. B\"urger and O. Hahn, Acta Polymer. \textbf{49}, 61 (1998).
}

\bibitem{Lyubartsev:2009p2497}
A. Lyubartsev, A. Mirzoev, L. Chen, and A. Laaksonen, Faraday Discuss. \textbf{144}, 43 (2010).  

\bibitem{Hansen:2005p292}
J.-P. Hansen, C. Addison, and A. A. Louis, J. Phys.: Condens. Matter \textbf{17}, S3185 (2005).  

\bibitem{Clark:2010p2483}
A. J. Clark and M. G. Guenza, J. Chem. Phys. \textbf{132}, 044902 (2010).  

\bibitem{Flory:1950p2484} 
P. J. Flory and W. R. Krigbaum, J. Chem. Phys \textbf{18}, 1086 (1950).  

\bibitem{Grosberg:1982p2265}
A. Y. Grosberg, P. G. Khalatur, and A. R. Khokhlov, Makromol. Chem., Rapid Commun. \textbf{3}, 709 (1982).  

\bibitem{Dautenhahn:1994p2250}
J. Dautenhahn and C. K. Hall, Macromolecules \textbf{27}, 5399 (1994).  

\bibitem{Murat:1998p1980} 
M. Murat and K. Kremer, J. Chem. Phys. \textbf{108}, 4340 (1998).
  
\bibitem{Bolhuis:2001p268} 
P. G. Bolhuis, A. A. Louis, J. P. Hansen, and E. J. Meijer, J. Chem. Phys. \textbf{114}, 4296 (2001). 
  
\bibitem{Pelissetto:2005p296}
A. Pelissetto and J.-P. Hansen, J. Chem. Phys. \textbf{122}, 134904 (2005).

\1{
\bibitem{D'Adamo2012}
G. D'Adamo, A. Pelissetto and C. Pierleoni, ``Polymers as soft compressible spheres'', J. Chem. Phys. in press (2012), http://arxiv.org/abs/1205.5654.
}  

\bibitem{Coluzza2008}I. Coluzza and J.-P. Hansen, Phys. Rev. Lett. \textbf{100}, 016104 (2008).
  
\bibitem{Pierleoni:2006p159}
C. Pierleoni, C. I. Addison, J. P. Hansen, and V. Krakoviack, Phys. Rev. Lett.  \textbf{96}, 128302 (2006).
  
\bibitem{HansenJP:2006p2248}
J.-P. Hansen and C. Pearson, Mol. Phys. \textbf{104}, 3389 (2006).
  
\bibitem{Sambriski:2007p781}
E. Sambriski, M. Guenza, Phys. Rev. E \textbf{76}, 051801 (2007).
  
\bibitem{Capone2009}  B. Capone, C. Pierleoni, J.-P. Hansen, and V. Krakoviack, J.  Phys. Chem. B \textbf{113} , 3629 (2009).
  
\bibitem{Gross:2010p2528}
C. Gross,W. Paul, Soft Matter \textbf{6}, 3273 (2010).

\bibitem{Capone2010} I. Coluzza, B. Capone and J.-P. Hansen, Soft Matter \textbf{7}, 5255  (2011).
 
\bibitem{Capone2011} B. Capone, J.-P. Hansen and I. Coluzza,  J. of Phys.: Cond. Mat. \textbf{23}, 194102 (2011).

\bibitem{Doi}
M. Doi, \textit{Introduction to Polymer Physics} (Clarendon Press, Oxford, 1992).
  
\bibitem{Pierleoni:2007p193}
C. Pierleoni, B. Capone and J. P. Hansen, J. Chem. Phys. \textbf{127}, 171102 (2007).
  
\bibitem{Pelissetto:2009p287}
A. Pelissetto, J. Phys.: Condens. Matter \textbf{21}, 115108 (2009).

\1{
\bibitem{Vettorel:2010p1733} 
T. Vettorel, G. Besold, and K. Kremer, {Soft Matter\/} {\bf 6}, 2282 (2010).
}

\bibitem{DAdamoG:2011p2477}
G. D'Adamo, A. Pelissetto, and C. Pierleoni, \1{Soft Matter \textbf{8}, 5151 (2012).} 

\bibitem{Clisby:2010p2249}
N. Clisby, Phys. Rev. Lett. \textbf{104}, 55702 (2010).
 
\bibitem{Laso:1991p1417}
  M. Laso, H. {\"O}ttinger, and U. Suter, J. Chem. Phys. \textbf{95},
2178 (1991).
  
\bibitem{footnote}
To study polymer chains in the scaling limit, 
we adopted the Domb-Joyce lattice model with 
$w=0.5058$, the value for which the coefficient of the leading-order 
corrections to scaling approximatively vanishes \cite{Pelissetto:2008p1683}.
As discussed in the appendix of Ref.~\cite{DAdamoG:2011p2477}, we considered chains of lengths $L=600$ and 
2400 to extrapolate at the infinite chain length limit.

\bibitem{Pelissetto:2008p1683}
A. Pelissetto, J. Chem. Phys. \textbf{129}, 044901 (2008).

\bibitem{Caracciolo:2006p587}
S. Caracciolo, B. M. Mognetti, and A. Pelissetto, J. Chem. Phys. \textbf{125}, 094903 (2006).  

\bibitem{HansenMcDonald1987} 
J. P. Hansen and I. McDonald, \textit{Theory of Simple Liquids}, 3rd ed. (Academic Press, Amsterdam, 2006)
  
\bibitem{Akkermans:2001p1711}
R. L. C. Akkermans and W. J. Briels, J. Chem. Phys.  \textbf{114}, 1020  (2001).

\bibitem{Akkermans204} 
R. L. C. Akkermans and G. Ciccotti, J. Phys. Chem. B \textbf{108}, 6866 (2004).

\1{
\bibitem{Farnoux78} B. Farnoux, F. Boue, J. P. Cotton, M. Daoud, G. Jannink, M. Nierlich and P. G. de Gennes, J. Phys. (Paris) \textbf{39}, 77 (1978).

\bibitem{Paul91} W. Paul, K. Binder, D.W. Hermann and K. Kremer, J. Phys. II (Paris) ÿ\textbf{1}, 37 (1991).

\bibitem{Muller00} M. M\"uller, K. Binder and L. Sch\"afer, Macromolecules \textbf{33}, 4568 (2000).
}



\end{thebibliography}

\end{document}